\documentclass[aps,twocolumn,superscriptaddress]{revtex4}

\usepackage{graphicx}
\usepackage{amsmath}
\usepackage{amssymb}

\newcommand{\unites}[1]{\; \mathrm{#1}}   

\usepackage{multirow}

\begin{document}
\title{Inclusion of Coulomb effects in laser-atom interactions}

\author{J. Dubois}
\affiliation{Aix Marseille Univ, CNRS, Centrale Marseille, I2M, Marseille, France}
\author{S. A. Berman}
\affiliation{Aix Marseille Univ, CNRS, Centrale Marseille, I2M, Marseille, France}
\affiliation{School of Physics, Georgia Institute of Technology, Atlanta, Georgia 30332-0430, USA}
\author{C. Chandre}
\affiliation{Aix Marseille Univ, CNRS, Centrale Marseille, I2M, Marseille, France}
\author{T. Uzer}
\affiliation{School of Physics, Georgia Institute of Technology, Atlanta, Georgia 30332-0430, USA}

\begin{abstract}
We investigate the role of the Coulomb interaction in strong field processes. We find that the Coulomb field of the ion makes its presence known even in highly intense laser fields, in contrast to the assumptions of the strong field approximation. 
The dynamics of the electron after ionization is analyzed with four models for an arbitrary laser polarization: the Hamiltonian model in the dipole approximation, the strong field approximation, the Coulomb-corrected strong field approximation and the guiding center. 
These models illustrate clearly the Coulomb effects, in particular Coulomb focusing and Coulomb asymmetry. We show that the Coulomb-corrected strong field approximation and the guiding center are complementary, in the sense that the Coulomb-corrected strong field approximation describes well short time scale phenomena (shorter than a laser cycle) while the guiding center is well suited for describing long time scale phenomena (longer than a laser cycle) like Coulomb-driven recollisions and Rydberg state creation. 
\end{abstract}

\maketitle

\section{Introduction}
After ionization in an intense laser field, an ionized electron travels in the continuum until it reaches the detector, or it can come back to the ionic core and probe the target in a process called recollision. Recollisions are key processes in attosecond physics~\citep{Corkum1993, Becker2002, Corkum2007,  Agostini2008, Krausz2009, Becker2012}, since they are the origin of a variety of highly nonlinear phenomena, such as high harmonic generation (HHG), non-sequential double ionization (NSDI), and above-threshold ionization (ATI). Such processes are widely used in attosecond physics in order to obtain information about the target atoms or molecules. For example, it is possible to perform imaging of molecular orbitals~\citep{Meckel2008,Gaffney2007,Peng2015} and determine the electronic dynamics inside atoms or molecules~\citep{Blaga2012,Goulielmakis2010}. Historically, the first model of the recollision scenario makes use of the so-called strong field approximation (SFA), where the effects of the Coulomb interaction after tunnel-ionization are neglected~\citep{Keldysh1965}. The recollision scenario~\citep{Corkum1993,Schafer1993} based on the SFA has three steps: (i) The electron tunnel-ionizes through the barrier induced by the laser field on the ionic core potential~\citep{Keldysh1965,Ammosov1986}, (ii) travels in the laser field alone, and then upon return to the ionic core, (iii) recombines with the ion (and triggers HHG), or rescatters from the ionic core, either elastically (ATI) or inelastically (NSDI). 
\par
In step (ii) of the recollision scenario, the electrons are outside the ionic core region, in the continuum, and their dynamics is mainly classical. The main advantage of the use of the SFA in step (ii) is the analytic expressions of the trajectories [Eqs.~\eqref{eq:SFA_solutions}], which sheds some light on the recollision process. In some cases, the SFA encounters success, for instance for predicting the HHG cutoff~\citep{Corkum1993,Lewenstein1994} or frequency versus time profile of HHG radiation~\citep{Tate2007}. In other cases, the SFA is inaccurate when it is confronted with experimental results~\citep{Goreslavski2004,Landsman2013,Brabec1996}, especially when long time scale trajectories (of the order of multiple laser periods) are involved in the experiment. In linearly-polarized (LP) fields, the SFA suggests that if the electron does not return to the ionic core in less than one laser cycle after ionization, it never comes back to the core. However, recollisions involving multiple laser cycles have a significant effect in NSDI. Including the Coulomb interaction increases the NSDI probability (i.e., the recollision probability) by about one order of magnitude~\citep{Brabec1996, Bhardwaj2001, Yudin2001_1, Fu2001}. Coulomb effects play a significant role in ATI experiments as well~\citep{Comtois2005,Goreslavski2004, Landsman2013}. 
\par
Here we investigate the role of the Coulomb interaction in the recollision and ionization processes by shuttling between four models, namely: the Hamiltonian in the dipole approximation (referred to as the reference Hamiltonian), the SFA, the Coulomb-corrected SFA~\citep{Goreslavski2004} (CCSFA) and the guiding-center (GC) model~\citep{Dubois2018,Dubois2018_PRE}. The reference Hamiltonian combines both the laser and the Coulomb interaction. The SFA, which ignores the Coulomb interaction, is used to point out 	the contributions of the Coulomb potential in the reference model. The CCSFA and GC models decouple the laser and Coulomb interactions through perturbation theory and averaging, respectively, and they are used to analyze each interactions' contributions independently. We show that the CCSFA and GC models are complementary, in the sense that CCSFA describes well short time scale phenomena (shorter than a laser cycle) while the GC is well suited for describing long time scale phenomena (longer than a laser cycle) like the multiple laser cycle recollisions, which we refer to as Coulomb-driven recollisions. 
\par
In particular, in photoelectron momentum distributions (PMDs), there is an asymmetry with respect to the minor polarization axis, known as the Coulomb asymmetry~\citep{Goreslavski2004}, and a decrease of the final electron energy, referred to as Coulomb focusing~\citep{Brabec1996,Berman2015}. In Ref.~\citep{Dubois2018}, we introduced the GC model for the motion of ionized electrons, and we used it to identify the mechanism behind the bifurcation in the peak of the PMDs as a function of the ellipticity observed in experiments~\citep{Landsman2013,Li2017}. This bifurcation was attributed as a clear signature of the Coulomb effects. There, we also showed that the mechanism behind the bifurcation is closely related to Rydberg state creation~\citep{Nubbemeyer2008}, a process that cannot be described without the Coulomb interaction. In this article, on the one hand we show that the GC model can also be used to quantify the amount of Rydberg states creation and to demonstrate their close relation with the Coulomb-driven recollisions. On the other hand, we show that the Coulomb interaction always makes its presence known for long time scale phenomena such as ATI, in particular the Coulomb asymmetry, even at high intensity when the assumptions of the SFA are met. 
\par
The article is organized as follows: In Sec.~\ref{sec:models}, we describe step (i) coupled with step (ii) of the recollision scenario, using the reference Hamiltonian and the three reduced models (SFA, CCSFA and GC models) we employ throughout the article. In Sec.~\ref{sec:PMDs}, we analyze the PMDs and the initial conditions leading to the PMDs with the four models. In particular, we identify the domain of initial conditions leading to Rydberg state creation and Coulomb-driven recollisions, which we refer to as the rescattering domain. We find that the rescattering domain is absent in the SFA and the CCSFA, while it is present in the GC model. In Sec.~\ref{sec:Coulomb_driven_Rydberg_states}, we use the GC model to interpret Rydberg state creation and the mechanism behind Coulomb-driven recollisions. This model allows us to make qualitative predictions on these two phenomena, which we compare with classical trajectory Monte Carlo (CTMC) simulations. Finally in Sec.~\ref{sec:geometry_rescattering_domain}, we investigate the shape of the rescattering domain using the GC model. We show how the shape of the rescattering domain manifests itself in experiments, in particular, in the bifurcation of the PMDs~\citep{Dubois2018}. 

\section{The models \label{sec:models}}
First, we describe the reference model: The atom is described with a single active electron, the ionic core is set at the origin and is assumed to be static. The position of the electron is denoted $\mathbf{r}$, and its conjugate momentum is $\mathbf{p}$. In the length gauge~\citep{Peng2015} and the dipole approximation, the Hamiltonian governing the dynamics of an electron in an atom driven by a laser field reads
\begin{equation}
\label{eq:Hamiltonian_electron}
H (\mathbf{r}, \mathbf{p}, t) = \dfrac{|\mathbf{p}|^2}{2} + V(\mathbf{r})  + \mathbf{r} \cdot \mathbf{E} (t) ,
\end{equation}
where atomic units (a.u.) are used unless stated otherwise. We use the soft-Coulomb potential~\cite{Javanainen1988} to describe the ion-electron interaction, $V(\mathbf{r}) = - (|\mathbf{r}|^2 + 1)^{-1/2}$. The electric field is elliptically polarized,
\begin{equation*}
\mathbf{E} (t) = \dfrac{E_0 f(t)}{\sqrt{\xi^2+1}} \left[ \hat{\mathbf{x}}  \cos (\omega t) + \hat{\mathbf{y}} \xi \sin ( \omega t) \right]  .
\end{equation*}
The laser frequency we use is $\omega = 0.0584 \unites{a.u.}$ (corresponding to infrared light of wavelength $780 \unites{nm}$), the laser ellipticity is $\xi \in [0, 1]$, and the laser amplitude is $E_0 = 5.338 \times 10^{-9} \sqrt{I}$ with $I$ the intensity in $\mathrm{W}\cdot \mathrm{cm}^{-2}$. The laser envelope $f(t)$ is trapezoidal such that $f(t)=1$ for $t\in [0, T_p]$, $f(t) = (T_f - t)/(T_f - T_p)$ for $t \in [ T_p , T_f ]$ and $f(t) =0$ otherwise, where $T = 2\pi/\omega$ is the laser period. Here, $T_p$ and $T_f$ are the duration of the plateau and the laser pulse, respectively. Throughout the article, we use a two laser cycle ramp-down, i.e., $T_f = T_p + 2T$. The duration of the plateau is $T_p = 8T$ unless stated otherwise. The major and minor polarization axes are $\hat{\mathbf{x}}$ and $\hat{\mathbf{y}}$, respectively. 

\subsection{Step (i): Ionization model \label{sec:tunnel_ionization}}
When the laser field starts to oscillate, it creates an effective potential barrier through which the electron can tunnel-ionize. We use the Perelomov-Popov-Terent'ev~\citep{PerelomovI1966,PerelomovII1967,PerelomovIII1967} (PPT) theory to define the ionization rate and the initial conditions after ionization. The Keldysh parameter~\citep{Keldysh1965} $\gamma =  \omega\sqrt{2 I_p}/E_0$, where $I_p$ is the ionization potential, is used to estimate the dominant ionization process. If $\gamma \ll 1$, the ionization process is the adiabatic tunnel ionization~\citep{Keldysh1965,Ammosov1986}, i.e., the potential barrier is quasi-static during the tunneling. If $\gamma \gg 1$, the dominant process is multiphoton absorption. For $\gamma \sim 1$, the process is in between tunnel ionization and multiphoton absorption.
During this so-called nonadiabatic tunnel-ionization~\citep{PerelomovI1966,PerelomovII1967,PerelomovIII1967}, the wavepacket absorbs photons during the tunneling~\citep{Klaiber2015}. Here, we show a summary of part of the PPT theory~\citep{PerelomovI1966,PerelomovII1967,PerelomovIII1967} used in this article for the ionization rate and the initial conditions of the electron after ionization.
\par
The initial conditions and the ionization rate of the electron are parametrized by the ionization time $t_0$ and its momentum $\mathbf{p}_0$ at $t_0$. The ionization rate $W (t_0, \mathbf{p}_0)$ is given in Eq.~\eqref{app_eq:PPT_ionization_rate} while the initial conditions at $t = t_0$ of the electron are
\begin{subequations}
\label{eq:initial_conditions_tunneling}
\begin{eqnarray}
\mathbf{r}_0 &=& \dfrac{|\mathbf{E}(t_0)|}{\omega^2} (1 - \cosh \tau_0) \hat{\mathbf{n}}_{\parallel} (t_0)  ,  \\
\mathbf{p}_0 &=& p_{\parallel} \hat{\mathbf{n}}_{\parallel} (t_0)  + p_{\perp} \hat{\mathbf{n}}_{\perp} (t_0) + p_{z,0} \hat{\mathbf{z}} ,
\end{eqnarray}
\end{subequations}
with $\hat{\mathbf{n}}_{\parallel} (t_0) = \mathbf{E}(t_0)/ |\mathbf{E}(t_0)|$ and $\hat{\mathbf{n}}_{\perp} (t_0) = - [ \hat{\mathbf{n}}_{\parallel} (t_0) \cdot \hat{\mathbf{y}} ] \hat{\mathbf{x}} + [ \hat{\mathbf{n}}_{\parallel} (t_0) \cdot \hat{\mathbf{x}} ] \hat{\mathbf{y}}$. In other words, $p_{\parallel}$ is the initial momentum of the electron along the laser field direction, $p_{\perp}$ is the initial momentum transverse to the laser field direction in the polarization plane, and $p_{z,0}$ is the initial momentum perpendicular to the polarization plane $(\hat{\mathbf{x}},\hat{\mathbf{y}})$. The purely imaginary time $\tau_0 = \tau_0(t_0)$ is solution of the transcendental equation
\begin{equation}
\label{eq:transcandent_equation_tau}
\sinh^2 \tau_0 - \xi^2 \left( \cosh \tau_0 - \dfrac{\sinh\tau_0}{\tau_0} \right)^2 = \gamma_0(t_0)^2 ,
\end{equation}  
where $\gamma_0 (t_0) = \omega \sqrt{2 I_p}/|\mathbf{E}(t_0)|$. We consider $\mathrm{He}$ ($I_p = 0.9$) unless stated otherwise. According to the PPT ionization rate, the most probable trajectory ionizes at times $\omega t_0 = n \pi$ (see Appendix~\ref{app_sec:ionization_rate}), i.e., at the peak amplitude of the electric field, where $n\in\mathbb{N}$, with initial longitudinal, transverse and perpendicular momenta
\begin{subequations}
\label{eq:most_probable_initial_momentum}
\begin{eqnarray}
P_{\parallel} &=& 0 , \\
\label{eq:most_probable_initial_transverse_momentum}
P_{\perp} &=& \dfrac{\xi E_0}{\omega \sqrt{\xi^2+1}} \left( 1 - \dfrac{\sinh \tau}{\tau} \right) , \\
P_{z,0} &=& 0 ,
\end{eqnarray}
\end{subequations}
respectively, with $\tau$ the solution of Eq.~\eqref{eq:transcandent_equation_tau} for $\gamma_0(t_0) = \gamma \sqrt{\xi^2+1}$. Throughout the article, we consider $p_{z,0} = P_z = 0$ and, as a consequence, the electron dynamics is constrained to the polarization plane, i.e., $\mathbf{r}(t) \cdot \hat{\mathbf{z}} = 0$ and $\mathbf{p}(t) \cdot \hat{\mathbf{z}} = 0$ for all times $t$. We refer to the trajectory with initial conditions $(t_0= T/2, p_{\parallel} = P_{\parallel}, p_{\perp} = P_{\perp})$ as the T-trajectory. Hence, the ionization rate associated with the T-trajectory is the maximum ionization rate, and the T-trajectory captures the dominant behavior of the ionized electrons.

\subsection{Step (ii): Classical models}
\subsubsection{Reference Hamiltonian}
The reference Hamiltonian is defined in Eq.~\eqref{eq:Hamiltonian_electron} with initial conditions $(\mathbf{r}_0 , \mathbf{p}_0 , t_0)$ given by Eq.~\eqref{eq:initial_conditions_tunneling}. In order to derive the reduced models, we use a new set of phase-space coordinates $(\mathbf{r}_g,\mathbf{p}_g)$ such that
\begin{subequations}
\label{eq:Phi_2}
\begin{eqnarray}
\mathbf{r}_g &=& \mathbf{r} - \mathbf{\Sigma}(t)/\omega^2 , \\
\mathbf{p}_g &=& \mathbf{p} - \mathbf{A}(t) .
\end{eqnarray}
\end{subequations}
where $\omega^2 \mathbf{A}(t) = \partial \mathbf{\Sigma} (t)/\partial t$ and the vector potential $\mathbf{A}(t)$ is such that $\mathbf{E}(t) = - \partial \mathbf{A}(t)/\partial t$. Using integration by parts, $\mathbf{A}(t) = - f(t) E_0 [ \hat{\mathbf{x}} \sin ( \omega t) - \xi \hat{\mathbf{y}} \cos ( \omega t)]/\omega \sqrt{\xi^2+1} + O(T/T_f)$, where the terms of order $O(T/T_f)$ are due to the envelope variations. In the same way, we can show that $\mathbf{\Sigma}(t) = \mathbf{E}(t) + O(T/T_f)$. Here, we mainly focus on the analyses of the electronic dynamics during the plateau, and we consider $\mathbf{A}(t) \approx - f(t) E_0 [ \hat{\mathbf{x}} \sin ( \omega t) - \xi \hat{\mathbf{y}} \cos ( \omega t)]/\omega \sqrt{\xi^2+1}$ and $\mathbf{\Sigma}(t) \approx \mathbf{E}(t)$.
\par
Under the canonical change of coordinates~\eqref{eq:Phi_2}, Hamiltonian~\eqref{eq:Hamiltonian_electron} becomes 
\begin{equation}
\label{eq:guiding_center_model}
H_g (\mathbf{r}_g, \mathbf{p}_g, t ) = \dfrac{|\mathbf{p}_g|^2}{2} + V(\mathbf{r}_g + \mathbf{\Sigma}(t)/\omega^2) .
\end{equation}
The initial conditions in the new coordinates at time $t_0$, denoted $(\mathbf{r}_{g,0},\mathbf{p}_{g,0})$, are related to the old coordinates by the transformation~\eqref{eq:Phi_2} and such that 
\begin{subequations}
\label{eq:initial_conditions_guiding_center_coordinates}
\begin{eqnarray}
\mathbf{r}_{g,0} &=& \mathbf{r}_0 - \mathbf{\Sigma}(t_0)/\omega^2 , \\
\mathbf{p}_{g,0} &=& \mathbf{p}_0 - \mathbf{A}(t_0) . 
\end{eqnarray}
\end{subequations}
\par
The dynamics described by Hamiltonians~\eqref{eq:Hamiltonian_electron} and~\eqref{eq:guiding_center_model} are equivalent. In order to emphasize the role of the Coulomb interaction, we consider three reduced models in the new system of coordinates: the SFA where the Coulomb potential is neglected ($V = 0$), the CCSFA~\citep{Goreslavski2004} where the Coulomb potential is assumed to be a perturbation of the SFA prediction, and the GC model~\citep{Dubois2018,Dubois2018_PRE} in which the electron trajectory is averaged over one laser cycle. 

\subsubsection{SFA}
For the SFA and the CCSFA, we assume that the contribution of the Coulomb interaction on the electron dynamics acts on long time scales. Under this assumption, we write the Hamiltonian as
\begin{equation*}
H_g (\mathbf{r}_g, \mathbf{p}_g, t ) = \dfrac{|\mathbf{p}_g|^2}{2} + \epsilon V(\mathbf{r}_g + \mathbf{\Sigma}(t)/\omega^2) ,
\end{equation*}
where we have introduced an ordering parameter $\epsilon$ for bookkeeping purposes. We consider the correction due to the Coulomb interaction on a short time scale, hence $\mathbf{r}_g = \mathbf{r}^{\mathrm{SFA}}_g + \epsilon \Delta \mathbf{r}_g + O(\epsilon^2)$ and $\mathbf{p}_g = \mathbf{p}^{\mathrm{SFA}}_g + \epsilon \Delta \mathbf{p}_g + O(\epsilon^2)$. The lowest order in $\epsilon$ provides the SFA electron phase-space trajectory 
\begin{subequations}
\label{eq:SFA_solutions}
\begin{eqnarray}
\label{eq:SFA_solutions_r}
\mathbf{r}^{\mathrm{SFA}}_g (t) &=& \mathbf{r}_{g,0} + \mathbf{p}_{g,0} (t - t_0) , \\
\mathbf{p}^{\mathrm{SFA}}_g (t) &=& \mathbf{p}_{g,0} .
\end{eqnarray}
\end{subequations}
The electron trajectory, in this new set of coordinates, is that of a free particle with constant energy and drift momentum $|\mathbf{p}_g^{\mathrm{SFA}}|^2/2$ and $\mathbf{p}_g^{\mathrm{SFA}}$, respectively. At any time $t$, the position and momentum of the electron are obtained by inverting the change of coordinates~\eqref{eq:Phi_2}.

\subsubsection{Coulomb-corrected SFA}
The first order in $\epsilon$ provides the correction due to the Coulomb interaction on the SFA trajectory, which reads
\begin{subequations}
\label{eq:CCSFA_estimations}
\begin{eqnarray}
\Delta \mathbf{r}_g (t) &=&  \int_{t_0}^{t} \Delta \mathbf{p}_g (s) \; \mathrm{d}s , \\
\label{eq:CCSFA_estimations_Delta_p}
\Delta \mathbf{p}_g (t) &=& - \int_{t_0}^{t} \boldsymbol{\nabla} V\left( \mathbf{r}^{\mathrm{SFA}}_g(s) + \mathbf{\Sigma}(s)/\omega^2 \right) \; \mathrm{d}s .
\end{eqnarray} 
\end{subequations}
As mentioned above, the CCSFA is valid to determine the correction of the Coulomb interaction for short times (e.g., $t - t_0 \sim T$) regardless ellipticity. For longer times $t$, looking at Eqs.~\eqref{eq:SFA_solutions_r}, if the initial drift momentum of the electron $\mathbf{p}_{g,0}$ is large, the Coulomb correction~\eqref{eq:CCSFA_estimations_Delta_p} is significant only for a short time after ionization. According to the PPT theory, the initial drift momentum is of order $|\mathbf{p}_{g,0}| \sim \xi E_0/\omega$, hence, we expect the CCSFA to be valid only for large ellipticity.
\par
The integrals in Eqs.~\eqref{eq:CCSFA_estimations} are computed numerically. If the initial drift momentum of the T-trajectory is very large, the integrand in Eq.~\eqref{eq:CCSFA_estimations_Delta_p} is large for a very short time after ionization, so we make the approximation $\mathbf{E}(t) \approx \mathbf{E}(t_0) + \omega^2 (t-t_0) \mathbf{A}(t_0) - \omega^2 (t-t_0)^2 \mathbf{E}(t_0)/2$. As a consequence, the SFA trajectory~\eqref{eq:SFA_solutions_r} is quadratic in time. Taking $V(\mathbf{r}) \approx - 1/|\mathbf{r}|$ and the initial conditions of the T-trajectory to be such that $P_{\perp} (t-t_0) \ll |\mathbf{r}_0| + (t-t_0)^2 |\mathbf{E}(t_0)|/2$ (which becomes valid at high intensity) the correction of the asymptotic drift momentum is given by
\begin{equation}
\label{eq:approximation_T_trajectory_CCSFA}
\Delta \mathbf{p}_g \approx \dfrac{\pi \; \hat{\mathbf{n}}_{\parallel} (t_0)}{(2 |\mathbf{r}_0|)^{3/2} \sqrt{|\mathbf{E}(t_0)|}} - \dfrac{P_{\perp} \; \hat{\mathbf{n}}_{\perp} (t_0)}{2 |\mathbf{r}_0|^2 |\mathbf{E}(t_0)|} .
\end{equation}
In Ref.~\citep{Goreslavski2004}, a similar result is derived for $P_{\perp} = 0$. 

\subsubsection{Guiding-center model}
An alternative way to include the Coulomb interaction is to consider the averaged motion of the electron~\citep{Dubois2018,Dubois2018_PRE}. In Refs.~\citep{Dubois2018,Dubois2018_PRE}, we showed the electron trajectory can be viewed as oscillating around a GC trajectory with constant energy. Assuming that one laser cycle is short compared to the characteristic time of the ionized electron trajectory, the ordering parameter $\epsilon$ is such that the laser frequency is large, i.e., $\omega \mapsto \omega/\epsilon$ and the Hamiltonian may be written
\begin{equation}
\label{eq:Hamiltonian_guiding_center_epsilon}
H_g (\mathbf{r}_g, \mathbf{p}_g, t ) = \epsilon \left[ \dfrac{|\mathbf{p}_g|^2}{2} + V(\mathbf{r}_g + \epsilon^2 \mathbf{\Sigma}(t)/\omega^2) \right] .
\end{equation}
Averaging Hamiltonian~\eqref{eq:Hamiltonian_guiding_center_epsilon} over the fast time scale~\citep{Dubois2018,Dubois2018_PRE}, at the second order in the ordering parameter $\epsilon$, leads to
\begin{equation}
\label{eq:H_n}
\bar{H}_g (\bar{\mathbf{r}}_g , \bar{\mathbf{p}}_g) = \dfrac{|\bar{\mathbf{p}}_g|^2}{2} + V (\bar{\mathbf{r}}_g)  .
\end{equation}
The initial conditions of the GC trajectory are given by Eqs.~\eqref{eq:initial_conditions_guiding_center_coordinates}. The reconstructed electron trajectory is given by inverting Eqs.~\eqref{eq:Phi_2}, where $(\bar{\mathbf{r}}_g (t) , \bar{\mathbf{p}}_g (t))$ are trajectories of Hamiltonian~\eqref{eq:H_n}. Therefore, the electron oscillates around its GC motion described by Hamiltonian~\eqref{eq:H_n}. The GC Hamiltonian is invariant under time translation, implying that its energy $\mathcal{E} = \bar{H}_g (\bar{\mathbf{r}}_g, \bar{\mathbf{p}}_g)$ is conserved. In addition, for a rotationally invariant potential as in the case of atoms and in particular the soft Coulomb potential used in this article, the angular momentum $\mathbf{L} = \bar{\mathbf{r}}_g \times \bar{\mathbf{p}}_g$ is conserved. Hence, there are as many conserved quantities as degrees of freedom, and Hamiltonian~\eqref{eq:H_n} is integrable.
\par
By substituting $\tau_0 \approx \sinh^{-1} \gamma_0 (t_0)$ (which holds for all ellipticities if $\gamma \lesssim 5$, see Ref.~\citep{Mur2001}) and considering $|\mathbf{E} (t_0 )| \sim E_0$ in Eq.~\eqref{eq:initial_conditions_tunneling}, the typical distance between the electron and the ionic core after tunnel-ionization is $|\mathbf{r}_0| \sim (E_0/\omega^2) ( 1 - \sqrt{1+\gamma^2} )$. The GC model is quantitatively accurate when $|\mathbf{r}_0| \gtrsim E_0/\omega^2$ \citep{Dubois2018_PRE}, and as a consequence, we expect the GC model to be quantitatively accurate for $\gamma \gtrsim 1.6$. 

\section{Photoelectron momentum distributions (PMDs) \label{sec:PMDs}}
The laser-atom interaction gives rise to complex phenomena, involving multiple temporal and spatial scales. The phenomena arising from short time scale and long time scale processes manifest themselves in different aspects of the measurements and for different values of the parameters. In this section, we analyze the influence of short vs. long time scale microscopic phenomena on macroscopic measurements like the photoelectron momentum distributions in the light of the reduced models described in the previous section, and in particular the CCSFA and the GC models.

\subsection{Short time scale dynamics \label{sec:comparison_models_intensity}}

\begin{figure}
\includegraphics[width=.5\textwidth]{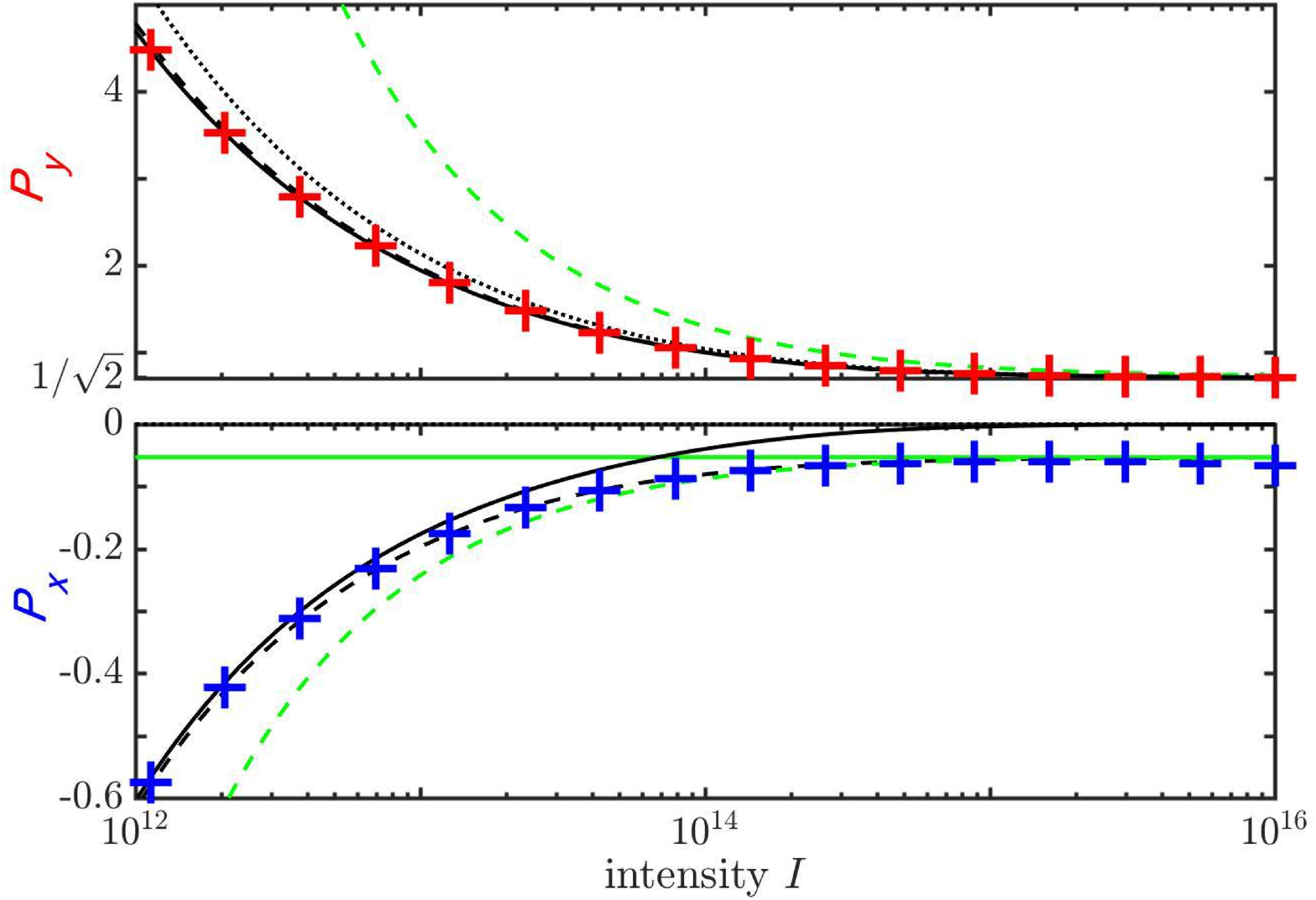}
\caption{T-trajectory final momentum $\mathbf{P} = P_x \hat{\mathbf{x}} + P_y \hat{\mathbf{y}}$ as function of the laser intensity $I$ for $\xi = 1$. The crosses are the T-trajectory final momentum of the reference Hamiltonian~\eqref{eq:Hamiltonian_electron}, where $P_x$ and $P_y$ are in blue and red, respectively. The dotted, dashed and solid black curves are the T-trajectory final momentum of the SFA, the CCSFA and the GC model, respectively. The dashed and solid green lines are the approximated and the asymptotic T-trajectory final momentum using Eqs.~\eqref{eq:approximation_T_trajectory_CCSFA} and~\eqref{eq:asymptotic_CCSFA_T_trajectory}, respectively. Momenta are scaled by $E_0/\omega$.}
\label{fig:Px_Py_intensity}
\end{figure}

First, we consider a circularly polarized (CP) field ($\xi = 1$), used for attoclock measurements~\citep{Torlina2015,Liu2017}. For ellipticity close to $1$, the initial drift momentum is large and the electron goes away from the ionic core quickly. Therefore, the corrections due to the Coulomb potential on the electron trajectories occurs on a short time scale, and we expect the CCSFA to be accurate. In attoclock measurements, the observable is the offset angle $\Theta$. We assume that it corresponds to the scattering angle of the T-trajectory
\begin{equation*}
\Theta = \tan^{-1}( P_y / P_x ) ,
\end{equation*}
where $\mathbf{P} = P_x \hat{\mathbf{x}} + P_y \hat{\mathbf{y}}$ is the T-trajectory final momentum (typical such trajectories are depicted in Fig.~\ref{fig:T_trajectory_analyses} for other parameters). Note that, to see the Coulomb asymmetry in a PMD from a CP field, a short laser pulse has to be used~\citep{Torlina2015,Liu2017}; otherwise, the PMD would resemble a ring around the origin. Figure~\ref{fig:Px_Py_intensity} shows the T-trajectory final momentum as a function of the intensity $I$ for $\xi = 1$. For $P_y$ (upper panel), we notice that the dashed black curves (CCSFA), the solid black curves (GC model) and the crosses [reference Hamiltonian~\eqref{eq:Hamiltonian_electron}] overlap for $I \in [10^{12},10^{16}] \; \mathrm{W} \cdot \mathrm{cm}^{-2}$, and hence a good agreement between these three models is observed. In addition, we notice that the value of $P_y$ predicted by these three models is lower compared to the SFA model. This is a microscopic (at the level of the trajectory) signature of the Coulomb focusing. Concerning the green dashed curve [which is the approximation of the CCSFA given by Eqs.~\eqref{eq:approximation_T_trajectory_CCSFA}], we observe that the approximation of the CCSFA [Eqs.~\eqref{eq:CCSFA_estimations}] becomes good only at high intensity $I \gtrsim 10^{15} \; \mathrm{W}\cdot\mathrm{cm}^{-2}$, where the drift momentum $|\mathbf{p}_{g,0}| \sim E_0/\omega$ is very large and where the electron spends a very short time close to the ionic core. At a very high intensity $I \sim 10^{16} \; \mathrm{W} \cdot \mathrm{cm}^{-2}$, all models converge to the same value predicted by the SFA $P_y^{\mathrm{SFA}} = ( E_0 / \omega ) /\sqrt{2}$.
\par
For $P_x$ (lower panel), we observe that the dashed black curves (CCSFA) and the crosses [Hamiltonian~\eqref{eq:Hamiltonian_electron}] overlap for $I \in [10^{12},10^{16}] \; \mathrm{W} \cdot \mathrm{cm}^{-2}$. The solid black curve (GC model) agrees well with the crosses [reference Hamiltonian~\eqref{eq:Hamiltonian_electron}] only for intensities such that $I \lesssim  8 \times 10^{13} \; \mathrm{W}\cdot \mathrm{cm}^{-2}$. This intensity range corresponds to a Keldysh parameter $\gamma \gtrsim 1.6$ for which the electron initial position is $|\mathbf{r}_0 | \gtrsim E_0/\omega^2$, i.e., for which the GC model is quantitatively accurate. When the electron ionizes close to the ionic core, there is a large contribution of the Coulomb potential. Mapping the electron coordinates to its GC coordinates [Eq.~\eqref{eq:Phi_2}], and evaluating the Coulomb interaction on its GC only [Hamiltonian~\eqref{eq:H_n}] leads to a significant underestimate of the Coulomb effect if the electron is initially close to the ionic core. In the CCSFA, the evaluation of the Coulomb potential is performed on the approximate solution of the SFA. As a consequence, on a short time scale after ionization, the evaluation of the Coulomb interaction is performed on a position which is close to the real trajectory [Hamiltonian~\eqref{eq:Hamiltonian_electron}] and therefore close to the core. 
\par
We also observe that the dotted curve (SFA, $P_x^{\mathrm{SFA}} = 0$) never agrees with the crosses [reference Hamiltonian~\eqref{eq:Hamiltonian_electron}], even at very high intensity. This is a microscopic signature of the Coulomb asymmetry. In particular, we observe that the Coulomb asymmetry persists even for high intensity. For $I \gtrsim 10^{15} \; \mathrm{W}\cdot \mathrm{cm}^{-2}$, we observe that the dashed green curve [Eq.~\eqref{eq:approximation_T_trajectory_CCSFA}] agrees well with the dashed black curve (CCSFA) and the crosses [reference Hamiltonian~\eqref{eq:Hamiltonian_electron}]. For very high intensities, or equivalently for very small Keldysh parameter, the correction to the T-trajectory final momentum using Eq.~\eqref{eq:approximation_T_trajectory_CCSFA} becomes
\begin{equation}
\label{eq:asymptotic_CCSFA_T_trajectory}
\lim_{E_0 \to \infty} \dfrac{\Delta \mathbf{p}_g}{(E_0/\omega)} = - \dfrac{\omega \pi \; \hat{\mathbf{x}}}{(2 I_p)^{3/2} \sqrt{\xi^2+1}} ,
\end{equation}
which is valid for high ellipticity. The offset angle measured in an attoclock experiment is asymptotically
\begin{equation*}
\lim_{E_0 \to \infty} \Theta = \pi - \tan^{-1} \dfrac{\xi (2 I_p)^{3/2}}{\omega \pi} .
\end{equation*} 
The larger the intensity, the closer to the core the electron is initiated, and thus the T-trajectory remains deflected by the ionic core. Consequently, the Coulomb asymmetry persists even at very high intensity. In addition, in the reference Hamiltonian~\eqref{eq:Hamiltonian_electron}, the larger the intensity, the larger the laser-atom interaction $\mathbf{r}\cdot\mathbf{E}(t)$ and the Coulomb potential contribution $V(\mathbf{r})$. Therefore, the competition between the Coulomb potential and laser interaction is always present even at high intensity, as shown in Ref.~\citep{Berman2015}.
\par
In summary, as expected for large ellipticities (i.e., close to CP), there is a very good agreement between the CCSFA [Eqs.~\eqref{eq:CCSFA_estimations}] and the reference model [Hamiltonian~\eqref{eq:Hamiltonian_electron}] for all intensities. Indeed, for large ellipticities, the electron initial drift momentum is also large, and the Coulomb potential acts significantly on the electron trajectory for a short time after ionization. The Coulomb interaction causes the deflection of the T-trajectory after ionization. For intensities $I \gtrsim 8 \times 10^{13} \; \mathrm{W}\cdot \mathrm{cm}^{-2}$, the GC model also captures this effect well. 

\subsection{Long time scale dynamics}
For lower ellipticities, we show that important properties of the system arising from long time scale processes, in particular  Coulomb-driven recollisions and Rydberg state creation, are well described by the GC model but not by the CCSFA. To illustrate this, we consider an intensity $I = 8 \times 10^{13} \; \mathrm{W}\cdot\mathrm{cm}^{-2}$ ($\gamma \sim 1.6$) and an ellipticity $\xi = 0.4$.

\subsubsection{Analysis of the ionized electron momentum}

\begin{figure*}
\centering
\includegraphics[width=\textwidth]{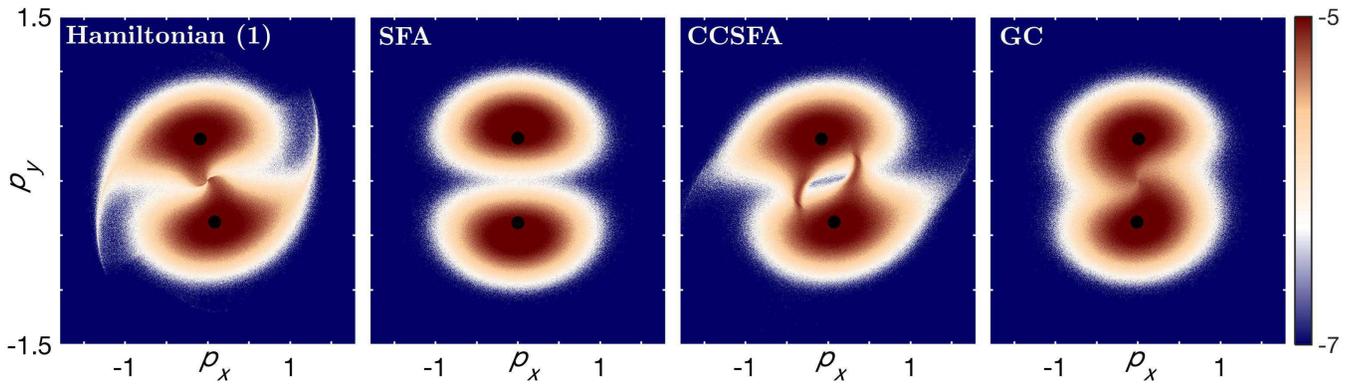}
\caption{Photoelectron momentum distributions (PMDs) in logarithmic scale for $I = 8\times 10^{13} \; \mathrm{W}\cdot \mathrm{cm}^{-2}$ and $\xi = 0.4$ of the reference Hamiltonian~\eqref{eq:Hamiltonian_electron}, the SFA~\citep{Corkum1993,Schafer1993}, the CCSFA~\citep{Goreslavski2004} and the GC model. The upper (resp. lower) black dot is the T-trajectory final momentum for each model (resp. its symmetric momentum with respect to the origin). The momenta are scaled by $E_0/\omega$.}
\label{fig:PMDs}
\end{figure*}

Figure~\ref{fig:PMDs} shows the PMDs computed with CTMC simulations of the reference Hamiltonian~\eqref{eq:Hamiltonian_electron}, the SFA [Eqs.~\eqref{eq:SFA_solutions}]~\citep{Corkum1993,Schafer1993}, the CCSFA [Eqs.~\eqref{eq:CCSFA_estimations}]~\citep{Goreslavski2004} and the GC model [Hamiltonian~\eqref{eq:H_n}]. The T-trajectory final momentum is shown with a black dot for each model. For each model, the PMDs are mainly two clouds centered around the T-trajectory final momentum. The two clouds are roughly symmetric with respect to the origin according to the symmetry $(\mathbf{r},\mathbf{p},t) \mapsto (-\mathbf{r},-\mathbf{p},t + T/2)$ of the reference Hamiltonian~\eqref{eq:Hamiltonian_electron} for a constant laser envelope ($f=1$) which is also preserved by the initial conditions [see Eq.~\eqref{eq:initial_conditions_tunneling}] and the reduced models. 
\par
For the reference Hamiltonian~\eqref{eq:Hamiltonian_electron} (leftmost panel of Fig.~\ref{fig:PMDs}), the PMD exhibits three significant features: The asymmetry with respect to the $\hat{\mathbf{y}}$-axis, the relatively high density of electrons with near-zero momentum --corresponding to near-zero energy photoelectrons~\citep{Xia2015}--, and the tails for high momentum (regions for $|p_x| > 1$). In order to interpret these features, we compare this PMD with those of the three reduced models.
\par
In the PMD of the SFA~\citep{Corkum1993,Schafer1993} (second panel of Fig.~\ref{fig:PMDs} from the left), the two clouds are symmetric with respect to the $\hat{\mathbf{y}}$-axis, there is a lack of near-zero energy photoelectrons, and there are no tails for high momentum. Therefore, these effects observed in the PMD of Hamiltonian~\eqref{eq:Hamiltonian_electron} are a consequence of the Coulomb potential, which is expected to be significant here since the characteristic time of the ionized trajectories is long compared to one laser cycle. 
\par
In the PMD of the CCSFA~\citep{Goreslavski2004} (third panel of Fig.~\ref{fig:PMDs} from the left), the two clouds are asymmetric with respect to the $\hat{\mathbf{y}}$-axis. As discussed in the previous section, after ionizing, the electron trajectories deviate because of the Coulomb interaction: This asymmetry is the Coulomb asymmetry. With the CCSFA, however, we observe that the distribution is very low around the origin of momentum space, i.e., there is still a lack of near-zero energy photoelectrons. Indeed, the drift momentum of the near-zero energy photoelectrons is low and the conditions on the validity of the CCSFA are not met. We notice that the integrals we compute numerically for determining the correction to the final momentum of the electron [Eqs.~\eqref{eq:CCSFA_estimations}] do not always converge. Obviously, the integrals diverge if for instance $\mathbf{p}_{g,0} = 0$. Also, for small drift momentum, it is challenging to obtain numerically converged integrals. Finally, we observe tails for $|p_x| > 1$ in the PMD of the CCSFA like in the PMD of Hamiltonian~\eqref{eq:Hamiltonian_electron}.
\par
In the PMD of the GC model (rightmost panel of Fig.~\ref{fig:PMDs}), the clouds are asymmetric with respect to the $\hat{\mathbf{y}}$-axis. After ionizing, the electron trajectories are deflected by the Coulomb force exerted on their GC~\citep{Dubois2018}. One advantage of this model is that the final momentum of the electron has an explicit expression for $V(\bar{\mathbf{r}}_g) \approx -1/|\bar{\mathbf{r}}_g|$ (see Appendix~\ref{app_sec:final_momentum_guiding_center}), and as a consequence the computations of the CTMCs are as fast as the computation of the CTMCs of the SFA. Moreover, this model does not rely on computing integrals that may or may not converge. In addition, we observe that the asymmetric clouds are connected to the origin of the momentum space, showing that the near-zero energy photoelectrons are well captured by this model. However, the absence of tails in the GC model suggests that the tails observed in the reference model and the CCSFA are the contribution of rescattered electrons~\citep{Kastner2012,Danek2018}.
\par
Hence, the asymmetry observed in the PMD of the reference Hamiltonian~\eqref{eq:Hamiltonian_electron} is also captured by the reduced models of the CCSFA and the GC. This asymmetry is due to the deviation of the electrons or their GC originating from the Coulomb interaction. In addition, near-zero-energy photoelectrons are captured by the GC model. The tails in the PMDs are due to the rescattering of electrons that have experienced soft recollisions~\citep{Kastner2012}, in which the electron comes close to the ionic core and is rescattered due to the competitive forces between the laser and the Coulomb interaction. This short time scale process is well known and well described by the CCSFA (see, e.g., Refs.~\citep{Kastner2012,Danek2018}). 

\subsubsection{Analysis of the initial conditions \label{sec:initial_conditions_analyses}}

\begin{figure}
\centering
\includegraphics[width=0.5\textwidth]{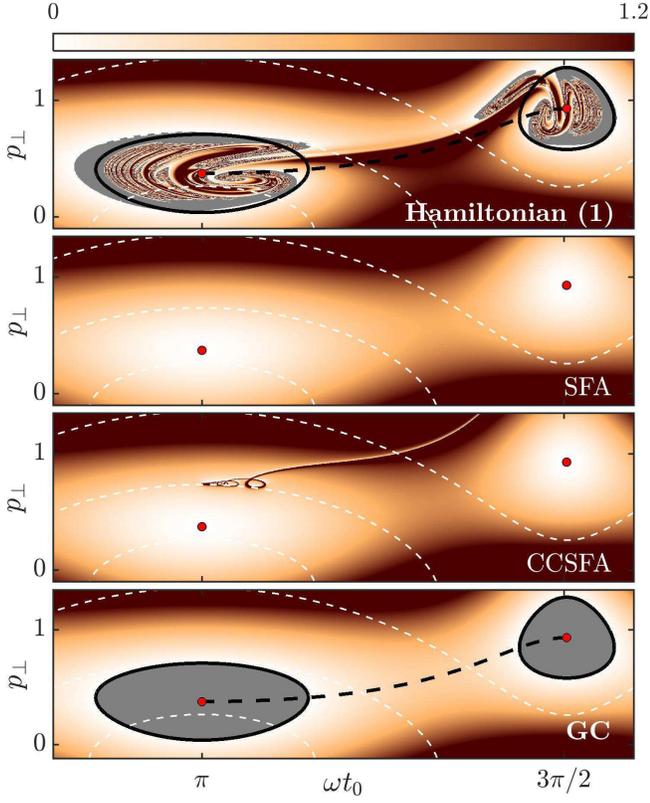}
\caption{Electron final energy as a function of the initial conditions $(t_0, p_{\perp}, p_{\parallel} = 0)$ for $I = 8\times 10^{13} \; \mathrm{W}\cdot \mathrm{cm}^{-2}$ and $\xi = 0.4$ of the reference Hamiltonian~\eqref{eq:Hamiltonian_electron}, the SFA~\citep{Corkum1993,Schafer1993}, the CCSFA~\citep{Goreslavski2004} and the GC model~\eqref{eq:H_n}. In grey-colored regions, the electron final energy is negative. The white dashed lines are  contours of constant ionization rate $W(t_0, p_{\perp})$ (see Appendix~\ref{app_sec:ionization_rate}) for $W / \max ( W ) =  10^{-1}$, $10^{-5}$ and $10^{-15}$, from bottom to top. The red dots correspond to the initial conditions for which the electron final energy in the SFA is zero, i.e., $\mathcal{E}^{\mathrm{SFA}} = 0$ [see Eq.~\eqref{eq:final_electron_energy_SFA}]. The solid black lines correspond to the initial conditions for which the GC energy is zero, i.e., $\mathcal{E} = 0$ [see Eq.~\eqref{eq:final_electron_energy_GC}]. The black dashed line corresponds to the initial conditions for which the GC angular momentum is zero and the initial radial momentum is negative, i.e., $\mathbf{L} = \boldsymbol{0}$ and $p_r (t_0) < 0$, where $\mathbf{L} = \bar{\mathbf{r}}_g \times \bar{\mathbf{p}}_g$ and $p_r = \bar{\mathbf{p}}_g \cdot \bar{\mathbf{r}}_g/|\bar{\mathbf{r}}_g|$. The momentum and the energy are scaled by $E_0/\omega$ and $\mathrm{U}_p = E_0^2/4\omega^2$, respectively.}
\label{fig:electron_final_energy}
\end{figure}

We investigate the initial conditions of the electron after ionization to interpret and understand the origin of the near-zero energy photoelectrons. Figure~\ref{fig:electron_final_energy} shows the final energy of the electron as a function of its initial conditions after tunneling for $I = 8\times 10^{13} \; \mathrm{W}\cdot \mathrm{cm}^{-2}$ and $\xi = 0.4$ for the reference Hamiltonian~\eqref{eq:Hamiltonian_electron}, the SFA [Eqs.~\eqref{eq:SFA_solutions}]~\citep{Corkum1993,Schafer1993}, the CCSFA [Eqs.~\eqref{eq:CCSFA_estimations}]~\citep{Goreslavski2004} and the GC model [Hamiltonian~\eqref{eq:H_n}]. The space of initial conditions is restricted to $p_{\parallel} = 0$, which is the most probable initial longitudinal momentum.
\par
For the reference Hamiltonian~\eqref{eq:Hamiltonian_electron} (upper panel of Fig.~\ref{fig:electron_final_energy}), we observe two grey regions of initial conditions where the electron final energy is negative, i.e., in which the electron is trapped in Rydberg states~\citep{Nubbemeyer2008}. The color corresponds to the final energy of photoelectrons which have reached the detector. Enveloped by the grey regions, we observe that there are ionized electrons whose energy depends extremely sensitively on the initial conditions, as a signature of the rescattering process. We refer to this region containing both the sensitivity to initial conditions and negative final energies as the rescattering domain. The boundaries of the rescattering domain are surrounded by regions of near-zero-energy photoelectrons. The part of the rescattering domain with small $p_{\perp}$ (lower part of the left grey regions) is in a region where the ionization rate is high. As a consequence, a significant number of electrons reach the detector with near-zero energy, as observed in the leftmost panel of Fig.~\ref{fig:PMDs}.
\par
For the SFA~\citep{Corkum1993,Schafer1993}, the final momentum of the electron is given by its initial drift momentum $\mathbf{p}_{g,0}$ since it is constant in time. As a consequence, the electron final energy is
\begin{equation}
\label{eq:final_electron_energy_SFA}
\mathcal{E}^{\mathrm{SFA}} = \dfrac{|\mathbf{p}_{g,0}|^2}{2} .
\end{equation} 
In the SFA (second panel from the top of Fig.~\ref{fig:electron_final_energy}), only two initial conditions lead to near zero-energy electrons, located at $p_{\perp} = -(E_0/\omega) \xi/\sqrt{\xi^2+1}$ and $\omega t_0 = \pi$, and at $p_{\perp} = -(E_0/\omega)/\sqrt{\xi^2+1}$ and $\omega t_0 = 3\pi/2$, represented by red dots in Fig~\ref{fig:electron_final_energy}. These initial conditions are located where the ionization rate is one or several orders of magnitude lower than the maximum ionization rate. The consequence is a lack of near-zero-energy photoelectrons in the PMD for the SFA observed in Fig.~\ref{fig:PMDs}. 
\par
For the CCSFA~\citep{Corkum1993,Schafer1993} (third panel from the top of Fig.~\ref{fig:electron_final_energy}), we observe the same patterns as for the SFA. The initial conditions of the near-zero-energy photoelectrons for the CCSFA are located in the same region of low ionization rate as for the SFA. Here again, the consequence is the lack of near-zero-energy electrons for the CCSFA~\citep{Goreslavski2004} observed in Fig.~\ref{fig:PMDs}. However, we observe in the CCSFA a region with an abrupt change of sensitivity to initial conditions across the light colored path, absent in the SFA. This path located at $p_{\perp} \approx 1$ at $\omega t_0 \in [\pi , 3\pi/2]$, in between the two red dots, separates near-zero-energy photoelectrons from high energy photoelectrons. This path is also present in the reference Hamiltonian at $p_{\perp} \approx 0.7$. It corresponds to soft recollisions~\citep{Kastner2012}.
\par 
The electron final energy using the GC model is given by
\begin{equation}
\label{eq:final_electron_energy_GC}
\mathcal{E} = \dfrac{|\mathbf{p}_{g,0}|^2}{2} + V \left( \mathbf{r}_{g,0} \right) .
\end{equation}
In the GC panel (lowest panel of Fig.~\ref{fig:electron_final_energy}), we observe the same region of initial conditions for which the electron final energy is negative as for the reference Hamiltonian~\eqref{eq:Hamiltonian_electron}, which corresponds to the rescattering domain.
The initial conditions for which the electron final energy is zero in the SFA are contained inside this region. The Coulomb potential creates this region in which the GC motion is bounded, which allows the electron to come back to the ionic core and to rescatter, or to be trapped into Rydberg states, scenarios analyzed in Sec.~\ref{sec:Coulomb_driven_Rydberg_states}. The boundaries of this rescattering domain correspond to the initial conditions for which the electron final energy is zero, i.e., $\mathcal{E} = 0$. We observe that the inclusion of the Coulomb potential pushes down the near-zero-energy photoelectrons to regions in momentum space for which the ionization rate is higher. As a consequence, we observe a significant number of near-zero energy photoelectrons in the PMD of the GC model. Moreover, we notice that $\mathcal{E} = \mathcal{E}^{\mathrm{SFA}} + V(\mathbf{r}_{g,0})$, and since the Coulomb potential is strictly negative, it is evident that electrons lose energy because of the Coulomb interaction, i.e., that electrons are subjected to Coulomb focusing.

\subsubsection{Types of trajectories}
 
\begin{figure}
\centering
\includegraphics[width=0.5\textwidth]{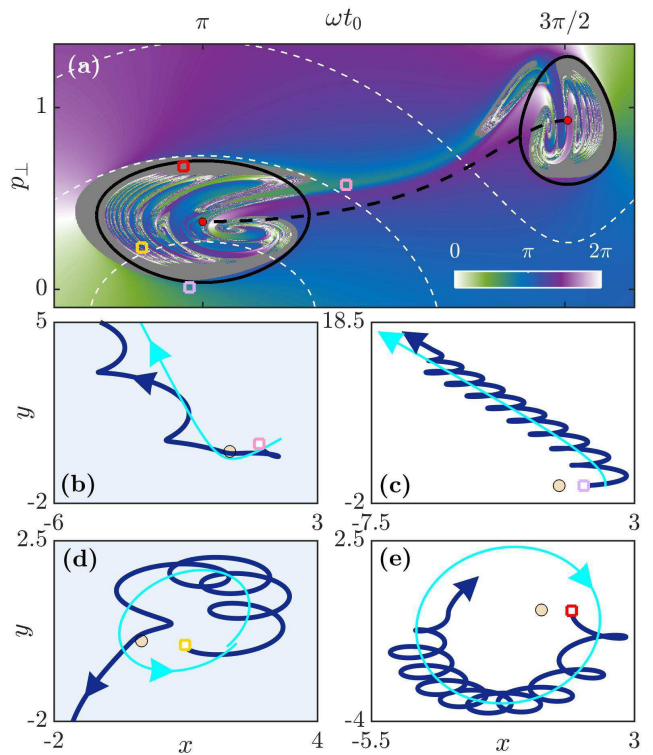}
\caption{(a) Scattering angle of the electron as a function of the initial conditions after tunneling $(t_0, p_{\perp}, p_{\parallel} = 0)$ for $I = 8\times 10^{13} \mathrm{W}\cdot \mathrm{cm}^{-2}$ and $\xi = 0.4$. The white dashed lines are the contours of constant ionization rate $W(t_0, p_{\perp})$ (see Appendix~\ref{app_sec:ionization_rate}) for $W / \max ( W ) =  10^{-1}$, $10^{-5}$ and $10^{-15}$, from bottom to top. The red dots correspond to the initial conditions for which $\mathcal{E}^{\mathrm{SFA}} = 0$ [see Eq.~\eqref{eq:final_electron_energy_SFA}]. The solid black line corresponds to the initial conditions for which $\mathcal{E} = 0$ [boundaries of the rescattering domain for the GC model, see Eq.~\eqref{eq:final_electron_energy_GC}]. The black dashed line corresponds to the initial conditions for which the GC angular momentum is zero and the initial radial momentum is negative, i.e., $\mathbf{L} = \boldsymbol{0}$ and $p_r (t_0) < 0$, where $\mathbf{L} = \bar{\mathbf{r}}_g \times \bar{\mathbf{p}}_g$ and $p_r = \bar{\mathbf{p}}_g \cdot \bar{\mathbf{r}}_g/|\bar{\mathbf{r}}_g|$. Grey areas show the conditions for which the electron is trapped into Rydberg states. (b--e) Dark and light blue curves are the electron and its GC trajectory, respectively. The initial condition of each trajectory is associated with a marker represented in (a). These trajectories represent a typical: (b) subcycle recollision, (c) direct ionization, (d) Coulomb-driven recollision, and (e) Rydberg state creation. Panels (b) and (c) have positive GC energy, while (d) and (e) have negative GC energy. Blue shaded panels indicate the cases with recollisions. The momentum and position are scaled by $E_0/\omega$ and $E_0/\omega^2$, respectively.}
\label{fig:tunneling_exit}
\end{figure}

In order to understand the origin of the sensitivity to initial conditions observed in the rescattering domain of the reference Hamiltonian~\eqref{eq:Hamiltonian_electron}, we analyze the different types of trajectories. Figure~\ref{fig:tunneling_exit}a shows the scattering angle of the electron, whose trajectory is obtained from the reference Hamiltonian~\eqref{eq:Hamiltonian_electron}, as a function of the initial conditions $(t_0 , p_{\perp})$ for $\xi = 0.4$. The scattering angle corresponds to the angle between the ionized electron momentum $\mathbf{p}$ at infinity and the major polarization axis ($\hat{\mathbf{x}}$-axis). In Figs.~\ref{fig:tunneling_exit}b--e, the dark blue curves are the electron trajectories of Hamiltonian~\eqref{eq:Hamiltonian_electron}, with initial conditions indicated by the corresponding markers in Fig.~\ref{fig:tunneling_exit}a. The light blue curves are the GC trajectories of Hamiltonian~\eqref{eq:H_n}. For Figs.~\ref{fig:tunneling_exit}c--e (as well for Fig.~\ref{fig:laser_pulse_duration}d--e and Fig.~\ref{fig:plateau_duration_trajectories}d), the GC is initialized far from the ionic core (for $|\mathbf{r} | \gtrsim 2 E_0/\omega^2$), during the plateau, in the domain of validity of the GC model (see Sec.~\ref{sec:T_trajectory_analyses} for a study of the discrepancy between the GC and the electron trajectory).
\par
Figure~\ref{fig:tunneling_exit}b shows a subcycle recollision. The initial condition of this trajectory is in the chaotic region near the condition for which the GC angular momentum is $\mathbf{L} = \bar{\mathbf{r}}_g \times \bar{\mathbf{p}}_g  \approx \boldsymbol{0}$ and the initial GC radial momentum is negative. Right after ionization, the GC trajectory is (mostly) straight, brings the electron to the core, and the electron recollides. The recollision occurs in a time scale shorter than one laser cycle, referred to as a subcycle recollision. We notice that if the electron tunnel-ionizes further away from the ionic core, the same conditions (near zero GC angular momentum) could lead to a multiple laser-cycle recollision.
\par
Figure~\ref{fig:tunneling_exit}c shows a direct ionization. The initial condition of this trajectory is in a regular region, for which the GC energy is positive $\mathcal{E} > 0$. The GC trajectory is unbounded, and leaves the ionic core region. The electron also leaves the ionic core region, driven by its GC.
\par
Figure~\ref{fig:tunneling_exit}d shows a Coulomb-driven recollision. The initial condition of this trajectory is in one of the main chaotic regions, for which the GC energy is negative $\mathcal{E} < 0$. The GC trajectory is bounded. As a consequence, the electron returns to the ionic core, driven by its GC, and recollides with the ionic core. After rescattering, the GC energy jumps to another energy level~\citep{Dubois2018_PRE}.
\par
Figure~\ref{fig:tunneling_exit}e shows a Rydberg state creation. The initial condition of this trajectory is in the grey area, for which the GC energy is negative $\mathcal{E} < 0$. The GC trajectory is bounded. However, contrary to the Coulomb-driven recollision (Fig.~\ref{fig:tunneling_exit}d), the laser pulse ends before the occurrence of the recollision. The Rydberg state creation corresponds to a frustrated Coulomb-driven recollision. The laser pulse duration plays an important role in determining the ratio between Coulomb-driven recollisions and Rydberg state trapping (see Sec.~\ref{sec:Coulomb_driven_recollision_laser_pulse_influence}).
\par
We observe that, in the four types of trajectories, two of them cannot be predicted by the SFA. While direct ionization and one-laser-cycle rescattering (Figs.~\ref{fig:tunneling_exit}c and~\ref{fig:tunneling_exit}b, respectively) are, at least qualitatively, predictable by the SFA, Coulomb-driven recollisions and Rydberg state creation (Figs.~\ref{fig:tunneling_exit}d and~\ref{fig:tunneling_exit}e, respectively) are predictable only when the Coulomb potential is taken into account. In the next section, we analyze Coulomb-driven recollisions and Rydberg state creation in more details.

\section{Coulomb-driven recollisions and Rydberg state creation \label{sec:Coulomb_driven_Rydberg_states}}
\subsection{Ionization time dependence \label{sec:Coulomb_driven_recollision_laser_pulse_influence}}

\begin{figure}
\centering
\includegraphics[width=0.5\textwidth]{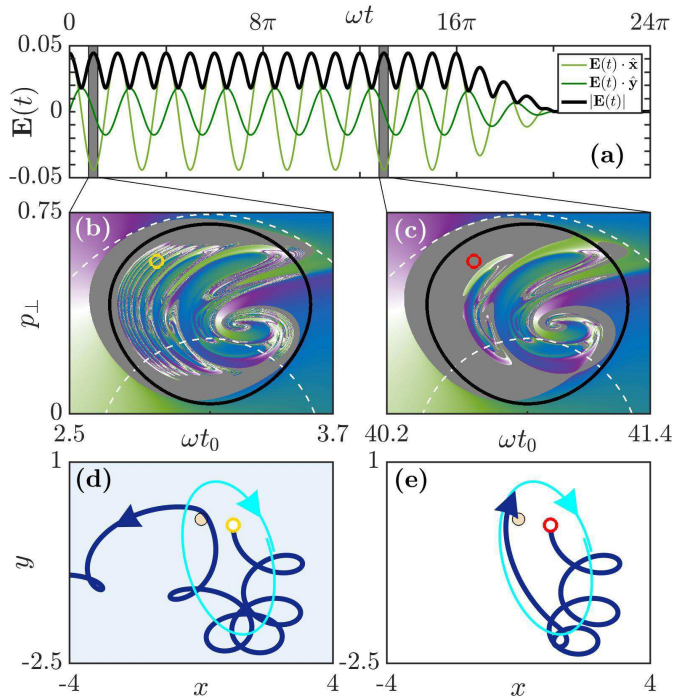}
\caption{(a) Electric field components and amplitude as a function of $\omega t$. The grey regions indicate the ionization time for which the final scattering angle is computed in (b) and (c). (b--c) Final scattering angle of the ionized electron as a function of the initial conditions $(\omega t_0, p_{\perp} , p_{\parallel} = 0)$ for $I = 8\times 10^{13} \; \mathrm{W}\cdot \mathrm{cm}^{-2}$ and $\xi = 0.4$ for the reference Hamiltonian~\eqref{eq:Hamiltonian_electron}.  The white dashed lines are the contour plot of the ionization rate $W(t_0, p_{\perp})$ (see Appendix~\ref{app_sec:ionization_rate}) for $W / \max ( W ) =  10^{-1}$, $10^{-5}$ and $10^{-15}$, from bottom to top. The solid black line corresponds to the initial conditions for which $\mathcal{E} = 0$ [boundaries of the rescattering domain for the GC model, see Eq.~\eqref{eq:final_electron_energy_GC}]. The dark grey region corresponds to the initial conditions for which the electron is trapped into a Rydberg state at the end of the pulse. (d--e) Dark and light blue curves are the electron and its GC trajectory, respectively. The initial conditions of the trajectories in (d) and (e) are indicated by circles in (b) and (c), respectively. The trajectories in (d) and (e) are initialized at the same laser phase, but (d) is a Coulomb-driven recollision and (e) is a Rydberg state creation. The momentum and the position are scaled by $E_0/\omega$ and $E_0/\omega^2$, respectively.}
\label{fig:laser_pulse_duration}
\end{figure}

Figures~\ref{fig:laser_pulse_duration}b and \ref{fig:laser_pulse_duration}c show the final scattering angle of the ionized electron as a function of its initial conditions $(\omega t_0, p_{\perp})$, for an ionization that takes place at the beginning of the pulse and at the end of the pulse, respectively (see Fig.~\ref{fig:laser_pulse_duration}a). Figures~\ref{fig:laser_pulse_duration}d and~\ref{fig:laser_pulse_duration}e show two electron trajectories with the same initial momentum and the same laser phase, but with two distinct ionization times (separated by six laser cycles). Since the phase is the same, the GC trajectories (light blue curves) in Figs.~\ref{fig:laser_pulse_duration}d and~\ref{fig:laser_pulse_duration}e are the same. Since the GC energy of these trajectories is negative, the GC trajectory is bounded. 
\par
In Fig.~\ref{fig:laser_pulse_duration}d, for which the electron ionizes at the beginning of the plateau, we observe that the electron oscillates around the bounded GC trajectory, which drives the electron back to the ionic core. After about four oscillations around the GC trajectory, the electron comes back to the ionic core. At this time, the GC energy jumps to another energy level due to the combined Coulomb and laser interaction, and the electron ionizes. This is a Coulomb-driven recollision.
\par 
In Fig.~\ref{fig:laser_pulse_duration}e, we observe that the electron oscillates as well around the bounded GC trajectory, which drives the electron back towards the core. However, when the electron is still far from the ionic core, the electric field is turned off, and the electron is trapped into a Rydberg state. The Rydberg state in which the electron is trapped corresponds almost to the Rydberg state of its GC.
\par
In other words, for both trajectories of Fig.~\ref{fig:laser_pulse_duration}d and~\ref{fig:laser_pulse_duration}e, the electron oscillates around the same GC trajectory. The difference between these two trajectories is the remaining time $10 T - t_0$ before the laser field is turned off. In Fig.~\ref{fig:laser_pulse_duration}d, the electron has enough time to undergo a close encounter with the ionic core ($|\mathbf{r}| < E_0/\omega^2$) before the electric field is turned off, when in Fig.~\ref{fig:laser_pulse_duration}e, the electric field turns off sooner, while the electron is still far from the ionic core ($|\mathbf{r}|> E_0/\omega^2$). The close encounter with the ionic core distinguishes the Coulomb-driven recollision from the Rydberg state creation. The scenarios of Coulomb-driven recollision and Rydberg state creation are closely related, since in both cases, the electron oscillates around a negative-energy GC. 

\begin{figure}
\includegraphics[width=0.5\textwidth]{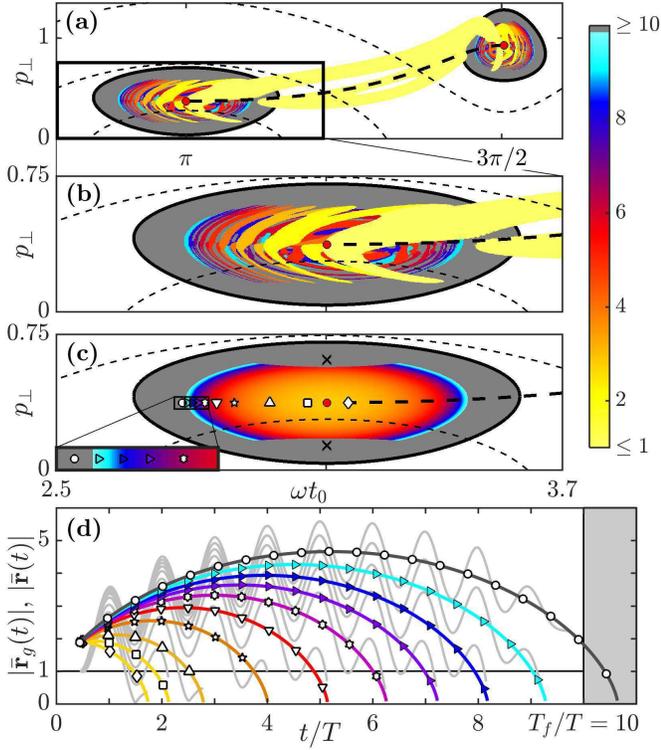}
\caption{Recollisions in the GC model for $I = 8 \times 10^{13} \; \mathrm{W} \cdot \mathrm{cm}^{-2}$ and $\xi = 0.4$. (a) $\Delta t/T$ as a function of the initial conditions $(t_0, p_{\perp}, p_{\parallel} = 0)$, where $\Delta t$ is the smallest time interval such that $|\bar{\mathbf{r}} (t_0+\Delta t)| = 5\; \mathrm{a.u.}$, with $t_0 + \Delta t < T_f$ (laser pulse duration $T_f = 10 T$) and $\bar{\mathbf{r}} (t) = \bar{\mathbf{r}}_g (t) + \mathbf{\Sigma}(t)/\omega^2$ (reconstructed trajectory). The white and dark grey regions are where this condition is never met, and where the GC energy is positive (white region) and negative (grey region). (b) Zoom of (a) on the largest rescattering domain. (c) $T_g/T = \omega/(2|\mathcal{E}|)^{3/2}$. The dark grey region is where the GC perihelion [see Eq.~\eqref{eq_app:perihelion_aphelion_negative_E}] is greater than the quiver radius $E_0/\omega^2 \approx 14 \; \mathrm{a.u.}$, or where $t_0 + T_g  \geq T_f$. The inset is a zoom. (a--c) The red dots and the black thick dashed curves are the same as in Fig.~\ref{fig:electron_final_energy}a. (c) The crosses are the location of the GC circular orbits (see Sec.~\ref{sec:long_plateau_duration_limit}). The dark dashed lines are contours of constant ionization rate $W(t_0, p_{\perp})$ (see Appendix~\ref{app_sec:ionization_rate}) for $W / \max ( W ) =  10^{-1}$, $10^{-5}$ and $10^{-15}$, from bottom to top. (d) The lines with markers are the GC trajectories $| \bar{\mathbf{r}}_g (t) |$ with initial conditions plotted in panel (c) with the corresponding marker and color, and the light grey lines are the reconstructed trajectories $| \bar{\mathbf{r}} (t) |$. The light grey region is when the laser field is turned off. Momenta are scaled by $E_0/\omega$.}
\label{fig:CDR_map}
\end{figure}

The GC model is used to interpret the relation between the Rydberg state creation and the Coulomb-driven recollisions. However, the initial conditions inside the rescattering domain corresponding to electrons that undergo Coulomb-driven recollisions in Figs.~\ref{fig:electron_final_energy}a, \ref{fig:tunneling_exit}a and~\ref{fig:laser_pulse_duration}b--c are not visible in Fig.~\ref{fig:electron_final_energy}d since the GC model does not describe the rescattering process. In order to see if the GC model also predicts which initial conditions lead to recollisions, we look at the excursion time $\Delta t$, which corresponds to the time the reconstructed electron trajectory spends before entering the region $|\bar{\mathbf{r}} (t_0 + \Delta t) | \leq R$ before the end of the laser field, where $\bar{\mathbf{r}} (t) = \bar{\mathbf{r}}_g (t) + \mathbf{\Sigma}(t)/\omega^2$ [see Eq.~\eqref{eq:Phi_2}], and $R = 5 \; \mathrm{a.u.}$ is an adjustable threshold.
\par
Looking at the excursion time per laser cycle $\Delta t/T$ of the GC model depicted in Figs.~\ref{fig:CDR_map}a--b, we observe similar patterns as for the reference model~\eqref{eq:Hamiltonian_electron} in Fig.~\ref{fig:electron_final_energy}a, \ref{fig:tunneling_exit}a and~\ref{fig:laser_pulse_duration}b--c. We have checked that these patterns are robust with respect to the value of the threshold $R$. 
For $\omega t_0 \in [\pi, 3\pi/2]$, the initial GC radial momentum is negative $p_r (t_0) = \mathbf{p}_{g,0} \cdot \mathbf{r}_{g,0} /|\mathbf{r}_{g,0}| < 0$, and therefore, in Fig.~\ref{fig:CDR_map}a, we observe a yellow region for $\mathbf{L} \approx \boldsymbol{0}$ corresponding to trajectories that recollide in shorter than one laser cycle after ionization, such as the one depicted in Fig.~\ref{fig:tunneling_exit}b (subcycle recollision). 
In Fig.~\ref{fig:CDR_map}b, which is a zoom of Fig.~\ref{fig:CDR_map}a around the largest rescattering domain, there are recollisions with excursion times of multiple laser cycles $\Delta t > T$. These are trajectories that spend multiple laser cycles far from the origin before returning to the ionic core, such as the one depicted in Fig.~\ref{fig:tunneling_exit}d (Coulomb-driven recollision). Hence, the GC model predicts qualitatively the initial conditions leading to recollision. We observe that the the initial conditions associated with the Coulomb-driven recollisions are organized in layers, similar to the layers observed in Figs.~\ref{fig:electron_final_energy}a, \ref{fig:tunneling_exit}a and~\ref{fig:laser_pulse_duration}b--c. Each layer is associated with a range of $\Delta t/T$ around an integer number, where, for decreasing ionization time for $\omega t_0 < \pi$, $\Delta t/T$ associated with each layer increases.
\par
In order to picture roughly the conditions for which the Coulomb-driven recollisions occur, we consider the period of the GC orbit per laser cycle $T_g/T = \omega / (2 |\mathcal{E}|)^{3/2}$ (using $V(\bar{\mathbf{r}}_g) \approx - 1/|\bar{\mathbf{r}}_g|$). Figure~\ref{fig:CDR_map}c shows the GC orbit period per laser cycle $T_g/T = \omega / (2 |\mathcal{E} |)^{3/2}$ as a function of the initial conditions in the largest rescattering domain. The grey regions correspond to the regions where the GC perihelion [see Eq.~\eqref{eq_app:perihelion_aphelion_negative_E}]---the closest distance between the GC orbit and the ionic core---is greater than $E_0/\omega^2$ or where $t_0 + T_g > T_f$. Figure~\ref{fig:CDR_map}d shows the GC distance from the ionic core $|\bar{\mathbf{r}}_g(t)|$ as a function of time per laser cycle of a sample of initial conditions indicated with the markers in Fig.~\ref{fig:CDR_map}c, and the distance from the ionic core $|\bar{\mathbf{r}} (t)|$ of the corresponding reconstructed trajectories. We see that the color code associated with the GC orbit period $T_g$ agrees well with the color code associated with the excursion time $\Delta t$ in Fig.~\ref{fig:CDR_map}a. Indeed, in Fig.~\ref{fig:CDR_map}d, we observe that the larger the period of the GC orbit followed by the electron, the larger its excursion time. As a consequence, the GC orbit period $T_g$ is a good observable to estimate the excursion time of the electron $\Delta t$. In addition, the GC orbit period of the trajectory associated with the leftmost marker in Fig.~\ref{fig:CDR_map}c is such that $t_0 + T_g > T_f$. The electron does not undergo recollision and ends up trapped in a Rydberg state since it comes back to the ionic core after the end of the laser pulse. Therefore, electrons undergoing Coulomb-driven recollisions are typically driven by GC orbits such that $T_g < T_f - t_0$.
\par
In summary, the electron is likely to undergo a Coulomb-driven recollision if it oscillates around a GC with a negative energy $\mathcal{E} < 0$, a positive initial GC radial momentum $p_r (t_0) = \mathbf{p}_{g,0} \cdot \mathbf{r}_{g,0} / | \mathbf{r}_{g,0} | > 0$, a GC orbital period such that $T_g = 2\pi/(2 | \mathcal{E}|)^{3/2} < T_f - t_0$ and a GC perihelion smaller than the quiver radius. Notice that the condition that the perihelion of the GC orbit is smaller than the quiver radius is equivalent to $\mathbf{L} \approx \boldsymbol{0}$. As a consequence, all recollisions are likely driven by small absolute values of the GC angular momentum. In contrast, the electron is likely to be trapped in a Rydberg state if it oscillates around a GC with a negative energy $\mathcal{E} < 0$ and either an orbital period greater than the laser pulse duration $T_g > T_f - t_0$ or a perihelion greater than the quiver radius, i.e., a large GC angular momentum. In the next section, we show that this latter process is robust due to the existence of center-saddle periodic orbits which are weakly unstable.

\subsection{Long plateau durations \label{sec:long_plateau_duration_limit}}

\begin{figure}
\centering
\includegraphics[width=0.5\textwidth]{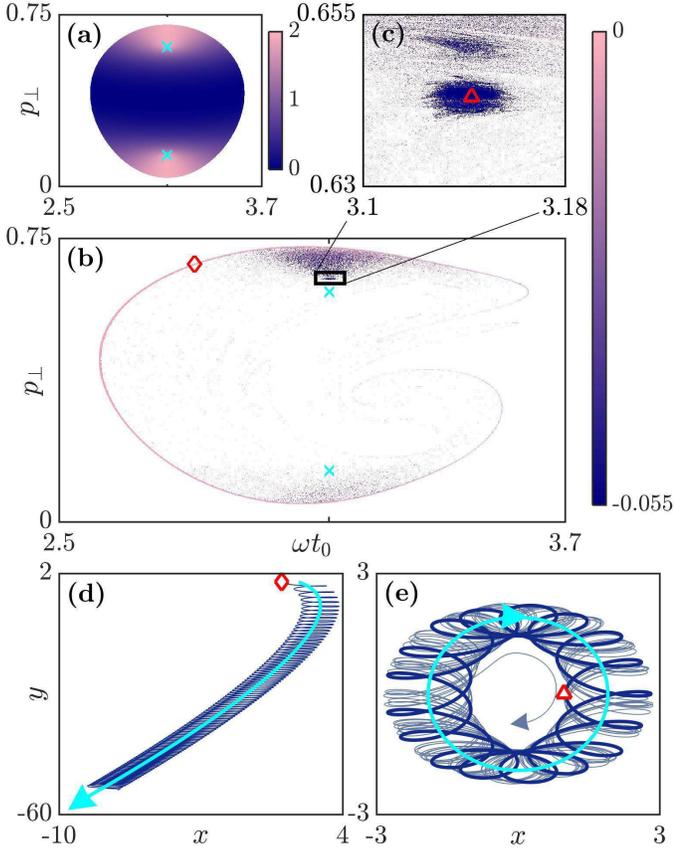}
\caption{The parameters are $I = 8\times 10^{13} \; \mathrm{W}\cdot\mathrm{cm}^{-2}$, $\xi = 0.4$ and plateau duration $T_p = 100 T$. (a) GC perihelion [see Eq.~\eqref{eq_app:perihelion_aphelion_negative_E}] in the rescattering domain depicted in the space of initial conditions $(t_0,p_{\perp},p_{\parallel})$. (b) Final negative energies of the electron trajectories of the reference Hamiltonian~\eqref{eq:Hamiltonian_electron}, where (c) is a zoom around the trapping region. The white color in (a--c) denotes an electron not trapped at the end of the laser pulse. (d),(e) Trajectories for the initial conditions indicated with a diamond in (c) and a triangle in (b), respectively. The dark blue and cyan curves are the electron trajectory of the reference Hamiltonian~\eqref{eq:Hamiltonian_electron} and the GC trajectory, respectively. The thick dark blue curve curve in (e) is a center-saddle periodic orbit very close to the region depicted in (c). The cyan crosses in (a) and (b) are the initial conditions of the clockwise (upper cross) and anticlockwise (lower cross) circular GC orbits. The cyan curve in (e) is the GC clockwise circular orbit, whose initial conditions are very close to the trapping region depicted in (c). Momenta and distance are scaled by $E_0/\omega$ and $E_0/\omega^2$, respectively.}
\label{fig:plateau_duration_trajectories}
\end{figure}

In Fig.~\ref{fig:CDR_map}a--b, we notice grey regions in the upper and lower part of the rescattering domain for which the GC orbit period is such that $T_g < T_f - t_0$. However, in these regions, the electron does not recollide because the GC perihelion is large (greater than $E_0/\omega^2 \approx 14 \; \mathrm{a.u.}$), as it is shown in Fig.~\ref{fig:plateau_duration_trajectories}a. As a consequence, there exists no time $\Delta t$ such that $|\bar{\mathbf{r}} (t_0+\Delta t)| $ is small, i.e., it is unlikely the electron recollides. This is also a scenario we observe in the reference model~\eqref{eq:Hamiltonian_electron}, in which the electron spins around the core for multiple laser cycles without recolliding. 
\par
For long plateau durations ($T_p = 100 T$, $T_f = T_p + 2T$) and an ionization time at the beginning of the laser pulse ($t_0 \ll T_p$), we expect that electrons oscillating around a negative near-zero energy GC (for which the GC orbit period is such that $T_g > T_f - t_0$) and electrons with a large GC perihelion [see Eq.~\eqref{eq_app:perihelion_aphelion_negative_E}] (GC perihelion greater than $E_0/\omega^2$ that prevents the electron from rescattering) create Rydberg states. In Fig.~\ref{fig:plateau_duration_trajectories}b, we observe indeed a pink thin layer of electrons creating Rydberg states, with a near-zero-energy GC such as the dark blue trajectory depicted in Fig.~\ref{fig:plateau_duration_trajectories}d. In addition, we observe two regions of initial conditions with smaller values of final energy for which the electrons are trapped in Rydberg states after having remained in the vicinity of the ionic core, for which the GC perihelion is larger than the quiver radius, as shown for the dark blue trajectory of the reference Hamiltonian in Fig.~\ref{fig:plateau_duration_trajectories}e. However, by comparing Fig.~\ref{fig:plateau_duration_trajectories}a and Fig.~\ref{fig:plateau_duration_trajectories}b, we observe that not all the electrons with a GC perihelion larger than the quiver radius are captured into Rydberg states. Here, we show how some electrons remain trapped while others do not.
\par
As observed in Fig.~\ref{fig:plateau_duration_trajectories}c, the filled region of initial conditions leading to electrons trapped in Rydberg states with a large GC perihelion is roughly regular. Figure~\ref{fig:plateau_duration_trajectories}e shows in dark blue a typical trajectory of Hamiltonian~\eqref{eq:Hamiltonian_electron} initiated inside this regular region. We observe that this trajectory turns around the core multiple times without being rescattered by the ionic core. As a consequence, the GC energy of this electron remains negative and roughly constant throughout the laser pulse duration~\citep{Dubois2018_PRE}. When the laser field is turned off, its GC energy is still negative and the electron is trapped in a Rydberg state. Near the initial conditions of this trajectory, there is a center-saddle periodic orbit of the reference model~\eqref{eq:Hamiltonian_electron} which exhibits the same pattern as this trajectory. This center-saddle periodic orbit is depicted in thick dark blue in Fig.~\ref{fig:plateau_duration_trajectories}e. In its neighborhood, the periodic orbit is center in one plane and saddle in a transverse plane defined by the eigenvectors of the monodromy matrix associated with the complex and real eigenvalues, respectively. Hence, there are two-dimensional invariant tori surrounding the periodic orbit in the center direction. The saddle direction is weakly unstable (its eigenvalue is $\sim 1.4$) and the orbit period is large (period of $30 T$), which implies that the unstable direction pushes slowly the electron away from each invariant torus. Consequently, trajectories in the vicinity of this periodic orbit remain close to it for relatively long times, even for long laser pulses. 
\par
In Fig.~\ref{fig:plateau_duration_trajectories}a, we observe that when the GC perihelion is large (greater than $E_0/\omega^2$), the recollisions are unlikely to happen as mentioned earlier. In these two regions of large GC perihelion, there are two cyan crosses indicating the initial conditions for which the GC orbit is circular. The initial conditions of these circular orbits are $p_{\perp} = \mathbf{A}(t_0)\cdot \hat{\mathbf{n}}(t_0) \pm \omega^2/|\mathbf{E}(t_0)|\cosh \tau_0(t_0)$ with $\omega t_0 = n\pi$ and $n\in\mathbb{N}$. They are close to the regular region in Figs.~\ref{fig:plateau_duration_trajectories}b--c. The circular orbit of the GC is depicted in cyan in Fig.~\ref{fig:plateau_duration_trajectories}e. We observe that the cyan curve provides the leading behavior of the averaged trajectory of the center-saddle periodic orbit in thick dark blue. The energy of the GC circular orbits (clockwise and anticlockwise) is given by $\mathcal{E} = - \omega^2 \sqrt{\xi^2+1} / (2 E_0 \cosh \tau)$ and their perihelion by $1/(2|\mathcal{E}|)$.
\par
In summary, there is a region of initial conditions for which the GC perihelion is larger than the quiver radius $E_0/\omega^2$, preventing the electron to recollide with the core. Instead, the electron is trapped in a Rydberg state. We showed that this process is robust because in the neighborhood of these initial conditions, there are center-saddle periodic orbits with weakly unstable directions that keep the electron in the vicinity of the core.

\subsection{Rate of Rydberg state creation \label{sec:rate_Rydberg_states}}

\begin{figure}
\centering
\includegraphics[width=0.5\textwidth]{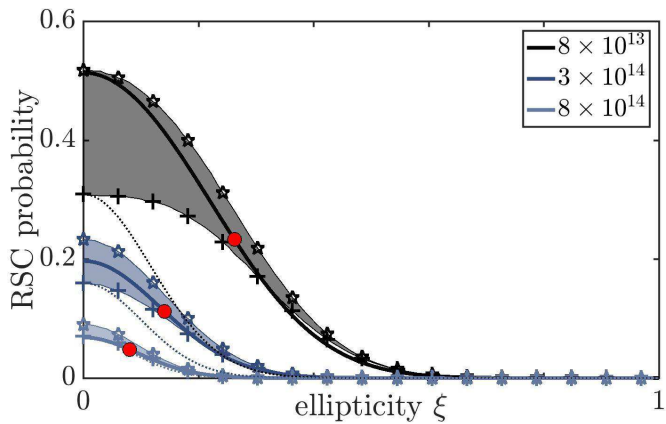}
\caption{Rydberg state creation (RSC) probability as a function of the laser ellipticity $\xi$ for $I = 8 \times 10^{13}$, $3\times 10^{14}$, and $8\times 10^{14}\;\mathrm{W}\cdot \mathrm{cm}^{-2}$. The RSC probability is defined as the ratio of the RSC yield $\mathbb{Y}$ to the ionized electron yield $N =  \int_{0}^{T_f} \mathrm{d}t_0 \int_{-\infty}^{\infty} \mathrm{d}^2 \mathbf{p}_0 \; W(t_0, \mathbf{p}_{0})$. The thick solid and dotted curves are our prediction with the GC model $\mathbb{Y}_{\mathrm{GC}} / N$ [Eq.~\eqref{eq:probabilty_rydberg_states_creation}], and the SFA $\mathbb{Y}_{\mathrm{SFA}} / N$, respectively. The thin curves with stars and crosses are the CTMC simulations of the reference Hamiltonian~\eqref{eq:Hamiltonian_electron} using $T_p = T$ and $T_p = 8T$, respectively. The filled areas show the estimate of the Coulomb-driven recollision probability for $T_p = 8T$. Red circles show the probability of RSC at the critical ellipticity $\xi_c$ given by Eq.~\eqref{eq:critical_ellipticity}.}
\label{fig:RSC_probability}
\end{figure}

Next, we investigate the rate of Rydberg state creation as a function of the laser ellipticity. A Rydberg state is created if the electron energy is negative at the end of the laser pulse. In the SFA, the condition of Rydberg state creation $\mathcal{E}^{\mathrm{SFA}} = 0$ [see Eq.~\eqref{eq:final_electron_energy_SFA}] is a one-dimensional curve $(t_0, \mathbf{p}^{\star}_0 (t_0))$ in a three-dimensional space $(t_0,p_{\parallel},p_{\perp})$, with $\mathbf{p}_0^{\star} (t_0) = \mathbf{A}(t_0)$. As a consequence, the probability of Rydberg state creation is in fact zero. In Refs.~\citep{Nubbemeyer2008,Landsman2013_NJP}, the yield of Rydberg state creation is given by $\mathbb{Y}_{\mathrm{SFA}} =  \int_{0}^{T_f} \mathrm{d}t_0 \; W(t_0, \mathbf{p}_{0}^{\star}(t_0))$.
\par
Figure~\ref{fig:RSC_probability} shows the Rydberg state creation probability as a function of the laser ellipticity from CTMC simulations of the reference Hamiltonian~\eqref{eq:Hamiltonian_electron} (thin solid curves with markers) and the SFA prediction (dotted curve) $\mathbb{Y}_{\mathrm{SFA}}/N$ with $N =  \int_{0}^{T_f} \mathrm{d}t_0 \int_{-\infty}^{\infty} \mathrm{d}^2 \mathbf{p}_0 \; W(t_0, \mathbf{p}_{0})$ the yield of ionized electrons. In Ref.~\citep{Landsman2013_NJP}, The SFA prediction is normalized such that it agrees at $\xi = 0$ with the CTMC simulations of the reference Hamiltonian~\eqref{eq:Hamiltonian_electron} for $T_p = 8T$ (thin lines with plus markers). Notice that only the SFA prediction is artificially normalized. For the SFA prediction (dotted curves), we observe a good agreement with the reference model at high ellipticity for all intensities and at low ellipticity for high intensity. However, there is a large discrepancy at low ellipticity for low and intermediate intensities, i.e., for $I \lesssim 5 \times 10^{14} \; \mathrm{W}\cdot \mathrm{cm}^{-2}$. For such intensities, the rescattering domain where Rydberg states arise is wide compared to the gradient of the ionization rate as observed in the top panel of Fig.~\ref{fig:electron_final_energy} and Fig.~\ref{fig:tunneling_exit}a. As a consequence, the SFA prediction that Rydberg states arise from the center of the rescattering domain is not accurate. 
\par
On the contrary, in Fig.~\ref{fig:electron_final_energy}, we see that the GC model is a good approximation for evaluating the size of the rescattering domain where the Rydberg states are created. In the GC model, a Rydberg state can be created only if the GC energy is negative $\mathcal{E} < 0$. As an approximation, we neglect the cases for which the electron undergoes a Coulomb-driven recollision according to the GC model. The GC prediction of the yield of Rydberg state creation is then given by 
\begin{equation}
\label{eq:yield_rydberg_states_creation_gc_pure}
\mathbb{Y}_{\mathrm{GC}} =  \int_{\Omega_R} W(t_0, \mathbf{p}_0) \; \mathrm{d}t_0 \mathrm{d}^2\mathbf{p}_0 ,
\end{equation}
where $\Omega_R = \lbrace t_0 \in [0,T_f] ,  \mathbf{p}_0 \in \mathbb{R}^2 \mid \mathcal{E} < 0 \rbrace$ is the set of initial conditions such that the GC energy is negative [see Eq.~\eqref{eq:final_electron_energy_GC}].
\par
According to Sec.~\ref{sec:Coulomb_driven_recollision_laser_pulse_influence}, an electron populating the rescattering domain either undergoes a Coulomb-driven recollision or is trapped in a Rydberg state. Hence, in order to minimize Coulomb-driven recollisions, we compare the GC prediction with CTMC simulations of the reference model for $T_p = T$. Figure~\ref{fig:RSC_probability} shows the GC prediction of Rydberg state creation probability (dashed curves) $\mathbb{Y}_{\mathrm{GC}}/N$. We observe an excellent agreement between the results of the simulation of the reference model~\eqref{eq:Hamiltonian_electron} for $T_p = T$ and the GC prediction for all ellipticities and intensities plotted here. For increasing intensity, the volume of the rescattering domain decreases, as shown in the next section. Hence, at high intensity, the ionization rate varies on large scales compared to the size of the rescattering domain, and the ionization rate is almost constant in the rescattering domain. Therefore, for high intensity, $\mathbb{Y}_{\mathrm{SFA}} \propto \mathbb{Y}_{\mathrm{GC}}$ as we observe in Fig.~\ref{fig:RSC_probability} for $I = 8\times 10^{14} \; \mathrm{W}\cdot \mathrm{cm}^{-2}$.

\section{The shape of the rescattering domain and its experimental implications \label{sec:geometry_rescattering_domain}}
\subsection{Analysis of the shape of the rescattering domain}

After ionization, the GC energy of the electron is given by Eq.~\eqref{eq:final_electron_energy_GC}. Substituting $\mathbf{A}(t_0) = [\mathbf{A}(t_0) \cdot \hat{\mathbf{n}}_{\parallel} (t_0)] \hat{\mathbf{n}}_{\parallel} (t_0) + [\mathbf{A}(t_0)\cdot \hat{\mathbf{n}}_{\perp} (t_0)]\hat{\mathbf{n}}_{\perp} (t_0)$ in Eq.~\eqref{eq:final_electron_energy_GC}, the rescattering domain defined by $\mathcal{E} < 0$ is the ensemble of initial conditions $(t_0, \mathbf{p}_0)$ such that 
\begin{equation}
\label{eq:rescattering_domain_given_t0}
\left[ p_{\parallel} - p_{\parallel}^{\star} (t_0) \right]^2 + \left[ p_{\perp} - p_{\perp}^{\star} (t_0) \right]^2  < \Delta p (t_0)^2 ,
\end{equation}
where $\Delta p (t_0) = \sqrt{2 | V(\bar{\mathbf{r}}_{g}(t_0)) | }$, $\mathbf{p}^{\star}_0 (t_0) = p_{\parallel}^{\star} (t_0) \hat{\mathbf{n}}_{\parallel} (t_0) + p_{\perp}^{\star} (t_0) \hat{\mathbf{n}}_{\perp} (t_0)$, hence $p_{\parallel}^{\star} (t_0) = \mathbf{A}(t_0) \cdot \hat{\mathbf{n}}_{\parallel}(t_0)$ and $p_{\perp}^{\star} (t_0) = \mathbf{A}(t_0) \cdot \hat{\mathbf{n}}_{\perp}(t_0)$. Here, $\bar{\mathbf{r}}_g(t_0) = - \mathbf{E}(t_0) \cosh \tau_0 (t_0) /\omega^2$ [see Eqs.~\eqref{eq:initial_conditions_guiding_center_coordinates}]. For a given ionization time $t_0$, the rescattering domain is a circle centered at $\mathbf{p}_0^{\star} (t_0)$ and of radius $\Delta p (t_0)$ in momentum space. The yield of Rydberg state creation in the GC model [see Eq.~\eqref{eq:yield_rydberg_states_creation_gc_pure}] becomes
\begin{widetext}
\begin{equation}
\label{eq:probabilty_rydberg_states_creation}
\mathbb{Y}_{\mathrm{GC}} =  \int_{0}^{T_f} \mathrm{d}t_0 \; \Delta p(t_0)^2 \int_0^1 \mathrm{d}\rho \; \rho \int_0^{2 \pi} \mathrm{d}\theta  \; W(t_0 , \mathbf{p}^{\star}_0 (t_0) + \rho \Delta p (t_0) \hat{\mathbf{n}} (t_0,\theta) ) ,
\end{equation}
\end{widetext}
where $\hat{\mathbf{n}}(t_0,\theta) = \hat{\mathbf{n}}_{\parallel} (t_0) \cos \theta + \hat{\mathbf{n}}_{\perp} (t_0) \sin\theta$. Equation~\eqref{eq:probabilty_rydberg_states_creation} is used to compute the yield of Rydberg state creation of Fig.~\ref{fig:RSC_probability}, and the integrals are performed numerically.
\par
\begin{figure}
\centering
\includegraphics[width=0.5\textwidth]{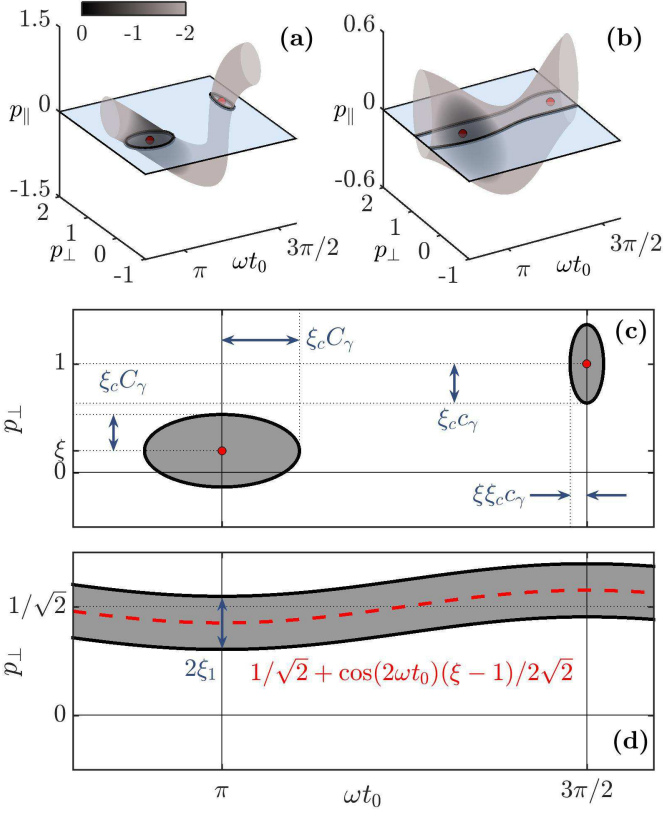}
\caption{Shape of the rescattering domain for $I = 8\times 10^{13} \; \mathrm{W}\cdot \mathrm{cm}^{-2}$ and $\xi = 0.2$ [close to LP, (a) and (c)], and $\xi = 0.7$ [close to CP, (b) and (d)]. (a), (b) Boundary of the rescattering domain as a function of the initial conditions $( t_0,p_{\parallel},p_{\perp})$. The color is the logarithm of the PPT ionization rate normalized by its maximum. The black lines are the boundaries of the rescattering domain for $p_{\parallel}=0$. (c),(d) Slice of the initial conditions $p_{\parallel} = 0$ [shaded blue planes in (a) and (b)]. Only the dominant orders in $\xi$ are depicted. Momenta are scaled by $E_0/\omega$.}
\label{fig:rescattering_domain_schematic}
\end{figure}

Figures~\ref{fig:rescattering_domain_schematic}a and \ref{fig:rescattering_domain_schematic}b show the boundaries of the rescattering domain in the space $(t_0,p_{\parallel},p_{\perp})$ for $\xi = 0.2$ and $\xi=0.7$. To see how the shape of the rescattering domain evolves as a function of the parameters, we focus on the conditions $p_{\parallel} =  0$ for which the ionization rate is maximum. For low ellipticity, the surface $p_{\parallel} = 0$ and the rescattering domain intersect in approximately ellipsoidal subdomains, while for high ellipticity, they intersect in a band. 

\subsubsection{Close to LP}
First, we consider the second order Taylor expansion of the shape of the rescattering domain for $p_{\parallel} = 0$ as a function of the ellipticity in the plane $(t_0, p_{\perp})$ close to LP ($\xi \ll 1$). For low ellipticity, the rescattering domain is approximately a set of ellipses, with two subsets: ellipses at the peak laser amplitude [around $\omega t_0 = n \pi$, with $n\in\mathbb{N}$], and ellipses at the lowest laser amplitude [around $\omega t_0 = (n+1/2)\pi$)]. 
\par
For $p_{\parallel} = 0$, the local minima of the final electron energy [see Eq.~\eqref{eq:final_electron_energy_GC}] are located at $\omega t_0^{\star} = n \pi / 2$ and $p_{\perp}^{\star} = p_{\perp}^{\star} (t_0^{\star})$ for $n \in \mathbb{N}$. The local minima of the final GC energy are the red dots depicted in Fig.~\ref{fig:rescattering_domain_schematic}c. In Eq.~\eqref{eq:rescattering_domain_given_t0}, we fix $p_{\parallel} = 0$ and we Taylor expand with respect to $t_0 - t_0^{\star}$. We obtain that the rescattering domain for $p_{\parallel} = 0$ can be written in the form
\begin{equation}
\label{eq:rescattering_domain_close_LP}
\dfrac{(p_{\perp} - p_{\perp}^{\star})^2}{\Delta p_{\perp}^2} + \dfrac{(t_0 - t_0^{\star})^2}{\Delta t_0^2} < 1 ,
\end{equation} 
where terms of order $(t_0 - t_0^{\star})^4$ and higher are neglected. Consequently, the subsets of rescattering domain in the plane $(t_0, p_{\perp})$ defined by $p_{\parallel} = 0$ are approximately ellipses and are centered around the local minima of the GC energy $( t_0^{\star}, p_{\perp}^{\star} )$. The expressions for $p_{\perp}^{\star}$, $\Delta p_{\perp}$ and $\Delta t_0$ depend on whether the ellipse is at the peak laser amplitude or at the lowest laser amplitude. 

\paragraph{Rescattering domains at the lowest laser amplitude:}
After Taylor expanding Eq.~\eqref{eq:rescattering_domain_given_t0} with respect to $t_0$ and $\xi$ around  the local minima $\omega t_0^{\star} = (n+1/2) \pi$ and $\xi = 0$, respectively, one gets (at the third order in the Taylor expansion) $p_{\perp}^{\star} \approx (E_0/\omega) (1 - \xi^2/2)$,
\begin{eqnarray*}
\Delta p_{\perp} &\approx& (E_0/\omega) \xi_c c_{\gamma} \left( 1 - \dfrac{\xi^2}{4 \gamma^2} \right) , \\
\omega \Delta t_0 &\approx& \xi \: \xi_c c_{\gamma} , 
\end{eqnarray*}
where $c_{\gamma} = \sqrt{\gamma} (1+\gamma^2)^{1/4} / \sinh^{-1}\gamma$, $\xi_c$ is defined in Eq.~\eqref{eq:critical_ellipticity}, and we have used $\tau \approx \sinh^{-1} \gamma$. Hence, at low ellipticities, the area of these ellipses is proportional to $\xi$ and consequently very small. For LP, the area of these ellipses is zero. In addition, at low ellipticities, these ellipses have a low weight given by the ionization rate, so their influence is negligible. 

\paragraph{Rescattering domains at the peak laser amplitude:}
After Taylor expanding Eq.~\eqref{eq:rescattering_domain_given_t0} with respect to $t_0$ and $\xi$ around  the local minima $\omega t_0^{\star} = n \pi$ and $\xi = 0$, respectively, one gets (at the third order in the Taylor expansion) $p_{\perp}^{\star} \approx \xi (E_0/\omega) (1 - \xi^2/2)$, 
\begin{subequations}
\label{eq:rescattering_domain_close_LP_expressions}
\begin{eqnarray}
\Delta p_{\perp} &\approx& (E_0/\omega) \xi_c C_{\gamma} \left[ 1 + \dfrac{\xi^2}{4 (1  + \gamma^2)} \right] , \\
\omega \Delta t_0 &\approx& \xi_c C_{\gamma} \left[ 1 + \xi^2 \dfrac{7 + 6 \gamma^2}{4 (1  + \gamma^2)} \right]  , 
\end{eqnarray}
\end{subequations}
where $C_{\gamma} = \gamma/\sinh^{-1} \gamma$ and we have used $\tau \approx \sinh^{-1} \gamma$. Here, the area of the ellipses is non-zero for LP, and because these ellipses are highly weighted by the ionization rate, they have a strong influence in the phenomena related to the rescattering domain such as, for instance, Rydberg state creation. We observe that for increasing intensity, these elliptical domains shrink towards their centers for which the GC energy is minimal (red dots in Fig.~\ref{fig:rescattering_domain_schematic}), which correspond also to the SFA conditions for which the electron final energy is zero [see Eq.~\eqref{eq:final_electron_energy_SFA}].

\subsubsection{Close to CP}
Next, we consider the second order Taylor expansion of the shape of the rescattering domain for $p_{\parallel} = 0$ as a function of the ellipticity in the plane $(t_0, p_{\perp})$ close to CP ($1 - |\xi| \ll 1$). For ellipticity close to $1$, the rescattering domain is approximately a band between two lines. We write Eq.~\eqref{eq:rescattering_domain_given_t0} in the form $p_{\perp} < p_{\perp}^{\star} (t_0) \pm [ p_{\parallel}^{\star} (t_0)^2 + \Delta p (t_0)^2 ]^{1/2}$. By Taylor expanding this expression to the first order (the second and third order expansion are too lengthy and do not provide additional relevant information to the discussion) with respect to $1 - |\xi|$ around $\xi = 1$, one gets that the lines surrounding the rescattering domain are
\begin{equation}
\label{eq:rescattering_domain_close_CP}
p_{\perp}^{\pm} (t_0) \approx \dfrac{E_0}{\sqrt{2}\omega} \left[ \cos ( 2 \omega t_0 ) \dfrac{\xi - 1}{2} + 1 \pm \xi_1 \right] , 
\end{equation} 
where $\xi_1 = (\omega^2 /E_0^{3/2}) (\gamma^2 + 1/2)^{-1/4}$ and we have used $\tau \approx \sinh^{-1} \gamma$. Hence, Coulomb-driven recollisions and Rydberg state creation after tunneling are likely when the lowest boundary line of the rescattering domain (see Fig.~\ref{fig:rescattering_domain_schematic}) approaches the regions of initial conditions with high ionization rate, i.e., $p_{\perp}^{-} (t_0) \lesssim P_{\perp}$. Fixing $\xi = 1$ and using Eq.~\eqref{eq:rescattering_domain_close_CP}, one gets
\begin{equation*}
E_0^{3/2} \lesssim \omega^2 \dfrac{\gamma (\gamma^2 + 1/2 )^{1/4}}{\sinh^{-1} \gamma} .
\end{equation*}
The term on the right-hand side of the inequality decreases for increasing $\gamma$. For $\gamma \ll 1$, the inequality becomes $I \lesssim 2 \times 10^{13} \; \mathrm{W} \cdot \mathrm{cm}^{-2}$. However, the condition $\gamma \ll 1$ implies that $I_p \ll 0.1 \; {\rm a.u.}$ in order for the electron to undergo a Coulomb-driven recollisions or be trapped in a Rydberg state at this frequency. Therefore, it is unlikely that the electron undergoes a Coulomb-driven recollision or is trapped in a Rydberg state for nearly-CP pulses, if the ionization takes place during the plateau. 

\subsection{Implication of the shape of the rescattering domain}

\begin{figure}
\centering
\includegraphics[width=0.5\textwidth]{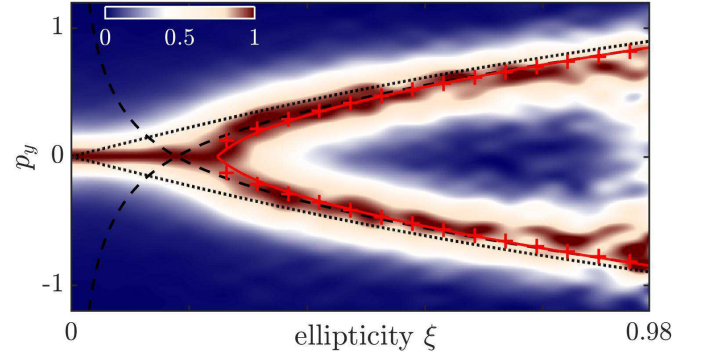}
\caption{Photoelectron momentum distribution along the minor polarization axis $\hat{\mathbf{y}}$ as a function of the ellipticity for $I = 1.2 \times 10^{14} \; \mathrm{W}\cdot \mathrm{cm}^{-2}$ and $\mathrm{Ar}$ ($I_p = 0.58 \; {\rm a.u.}$) and $\gamma \sim 1$. The color scale is the experimental data of Ref.~\citep{Li2017}. The dotted and dashed black lines are the T-trajectory of the SFA and the CCSFA, respectively. The cross markers and red solid lines are the T-trajectory of the reference Hamiltonian~\eqref{eq:Hamiltonian_electron} and the GC model~\eqref{eq:H_n}, respectively. Momenta are scaled by $E_0/\omega$.}
\label{fig:Bifurcations_1}
\end{figure}

In this section, we investigate the physical phenomena related to the shape of the rescattering domain, and we compare the results with experimental data. For instance, when the laser ellipiticity $\xi$ varies, the rescattering domain moves in phase space and as a consequence the PMDs change shape. In Fig.~\ref{fig:Bifurcations_1}, we show the experimental measurements from Ref.~\citep{Li2017}, of the final momentum distribution of the electron along the minor polarization axis $\hat{\mathbf{y}}$ as a function of the ellipticity $\xi$ for $I = 1.2\times 10^{14} \; \mathrm{W}\cdot \mathrm{cm}^{-2}$, $\mathrm{Ar}$ ($I_p = 0.58\; {\rm a.u.}$) and $\gamma \sim 1$. The experimental measurements of the final momentum along the minor polarization axis (color scale) show a distribution peaked around zero for small ellipticity. As the ellipticity increases, we observe a bifurcation of the peak of the distribution at a critical ellipticity $\xi_c \approx 0.25$, for which the distribution is no longer peaked around zero. After the bifurcation (for $\xi > \xi_c$), the peaks of the distribution move further apart for increasing ellipticity. 
\par
In Fig.~\ref{fig:Bifurcations_1}, we also show the $\hat{\mathbf{y}}$-component of the T-trajectory final momentum $P_y$ computed using the SFA $\mathbf{P}^{\mathrm{SFA}} = \hat{\mathbf{y}} \xi (E_0/\omega) \sinh\tau/(\tau\sqrt{\xi^2+1})$ (dotted lines), the CCSFA from Eq.~\eqref{eq:CCSFA_estimations} (dashed lines), the reference Hamiltonian~\eqref{eq:Hamiltonian_electron} (crosses) and the GC prediction [see Eq.~\eqref{eq:Final_momenta_guiding_center}] (solid lines). The prediction of the reference Hamiltonian~\eqref{eq:Hamiltonian_electron} is depicted only if the ionization is direct, i.e., if it has not undergone any recollisions and has not been trapped in Rydberg states. The GC prediction is depicted only when the GC energy is positive. Otherwise, the GC energy is negative and the electron does not reach the detector according to the GC model. We observe an excellent agreement between the experimental results from Ref.~\citep{Li2017}, the reference solution [Hamiltonian~\eqref{eq:Hamiltonian_electron}] and the GC prediction. 
\par
In a nutshell, for $\xi < \xi_c$, the T-trajectory is inside the rescattering domain. The GC motion is most often bounded, and as a consequence the electron undergoes recollisions or is trapped in a Rydberg state. When the ellipticity increases, the rescattering domain and the initial conditions of the T-trajectory move in phase space. At the critical ellipticity $\xi_c$, the T-trajectory is on the boundary of the rescattering domain, i.e., its GC energy is zero. For $\xi > \xi_c$, the GC motion is unbounded, and the electron ionizes without recollision. Therefore, the bunches in the PMDs after the bifurcation (as observed in Fig.~\ref{fig:PMDs}) are mainly composed of direct ionizations. Right after the bifurcation, a ridge structure can be seen for a certain range of laser parameters and atoms~\citep{Maurer2018,Danek2018}. The ridge structure is composed of near-zero-energy electrons that have undergone rescattering, and the bifurcation with ellipticity can be used to isolate these electrons from the electrons ionized directly~\citep{Maurer2018,Danek2018}. 

\subsubsection{Critical ionization time}
In LP fields, for $p_{\perp} = 0$ which reduces to a one-dimensional model, the SFA predicts that if an electron ionizes after a peak laser amplitude, i.e., at $t_0 > t_0^{\star}$ ($\omega t_0^{\star} = n \pi$ where $n\in\mathbb{N}$), it undergoes a recollision~\citep{Corkum1993}, while if it ionizes before this peak, i.e., at $t_0 < t_0^{\star}$, it ionizes directly. In the top panel of Fig.~\ref{fig:electron_final_energy} and in Figs.~\ref{fig:tunneling_exit}a and \ref{fig:laser_pulse_duration}a--b, we observe that this critical time $\omega t_0 = n \pi$ predicted by the SFA is lower if the Coulomb potential is taken into account, and according to the discussion in Sec.~\ref{sec:Coulomb_driven_recollision_laser_pulse_influence}, the electron potentially comes back to the ionic core even if it ionizes before the peak of the laser field.
\par
According to the GC model, using Eqs.~\eqref{eq:rescattering_domain_close_LP} and~\eqref{eq:rescattering_domain_close_LP_expressions} for $p_{\perp} = 0$, the left boundaries of the rescattering domain are given by $\omega t_c = n \pi - \xi_c C_{\gamma}$. If the electron ionizes at $t_0 < t_c$, the electron ionizes directly. If the electron ionizes at $t_0 > t_c$, the electron is in the rescattering domain. According to the discussion in Sec.~\ref{sec:Coulomb_driven_Rydberg_states}, the electron either populates Rydberg states, or undergoes a recollision. In particular, if an electron ionizes before the peak of the laser field and recollides, it is mainly because of the Coulomb interaction and the bounded motion of its GC that brings the electron back to the core. If the electron ionizes after the peak of the laser field, its GC initial radial momentum is negative (and its angular momentum is zero in 1D), and as a consequence the electron recollides. 
\par
The same arguments are extended to estimate $t_c$ for low ellipticity and $\xi \geq \xi_c$. We fix the initial momentum at its most probable value given by $(p_{\parallel} = P_{\parallel}, p_{\perp} = P_{\perp})$ and we let the ionization time $t_0$ free. At low ellipticity $P_{\parallel} \approx 0$ and $P_{\perp} \approx 0$, and if $\omega t_0 = \omega t_0^{\star}$ the trajectory is approximately at the center of the rescattering domain (see Fig.~\ref{fig:rescattering_domain_schematic}c). As a consequence, there exist intervals of ionization time $t_0$ for which the initial conditions are inside the rescattering domain, but also because of the shape of the rescattering domain (see Fig.~\ref{fig:rescattering_domain_schematic}c), there are intervals of ionization times $t_0$ for which the initial conditions are outside the rescattering domain. The critical time $t_c$ is the ionization time for which $(t_c , P_{\parallel}, P_{\perp})$ is on the boundary of the rescattering domain. In Eq.~\eqref{eq:rescattering_domain_close_LP}, we transform the inequality into an equality and we fix $p_{\perp} = P_{\perp} \approx \xi (E_0/\omega) ( 1 - \gamma/\sinh^{-1}\gamma)$ [see Eq.~\eqref{eq:most_probable_initial_transverse_momentum}]. Then, using Eqs.~\eqref{eq:rescattering_domain_close_LP_expressions} up to the second order in $\xi$, the critical time $t_c$ is given by
\begin{equation*}
\omega t_c \approx \omega t_0^{\star} - C_{\gamma} \sqrt{\xi_c^2 - \xi^2} , 
\end{equation*}
with $C_{\gamma} = \gamma/\sinh^{-1}\gamma$ and $\xi_c$ defined Eq.~\eqref{eq:critical_ellipticity}. Also, we have seen in Sec.~\ref{sec:Coulomb_driven_Rydberg_states} that if the ionization takes place after the peak of the laser field, the GC radial momentum is negative and the electron tends to recollide with the ionic core. Hence, the ionization time $t_0$ for direct ionization is $\omega t_0 \in \omega t_c - [ 0 , \pi/2 ]$. This is in agreement with the CTMC simulations of Ref.~\citep{Li2013}. 

\subsubsection{Critical ellipticity \label{sec:critical_ellipticity}}
Next, we consider the bifurcation with respect to the laser parameters. We consider the T-trajectory given by the initial conditions~\eqref{eq:most_probable_initial_momentum} and $\omega t_0 = \pi$. For LP ($\xi = 0$), the T-trajectory is inside the rescattering domain. As a consequence, the GC energy of the T-trajectory is negative, and the electron is either trapped in a Rydberg state, or undergoes a recollision. For increasing ellipticity, $p_{\perp}^{\star} (t_0)$ increases and $P_{\perp}$ decreases. At ellipticity $\xi = \xi_c$, the initial condition of the T-trajectory $(  \omega t_0 = \pi , p_{\parallel} = P_{\parallel} , p_{\perp} = P_{\perp} )$ crosses the boundary of the rescattering domain. In Eq.~\eqref{eq:rescattering_domain_close_LP}, we substitute the initial conditions of the T-trajectory $\omega t_0 = n \pi$ and $p_{\perp} = P_{\perp} \approx \xi (E_0/\omega) \gamma/\sinh^{-1}\gamma$ [see Eq.~\eqref{eq:most_probable_initial_transverse_momentum}] and we use Eq.~\eqref{eq:rescattering_domain_close_LP_expressions} up to the first order in $\xi$. Replacing $\xi$ by $\xi_c$ and assuming that $\xi_c^2 \ll 1$, the critical ellipticity is
\begin{equation}
\label{eq:critical_ellipticity}
\xi_c \approx \dfrac{\sqrt{2}\omega^2}{E_0^{3/2}} \dfrac{\sinh^{-1}\gamma}{\gamma (1+\gamma^2)^{1/4}} ,
\end{equation}
(see Supplemental Materials of Ref.~\citep{Dubois2018} for a detailed derivation). For $\xi > \xi_c$, the GC energy of the T-trajectory is positive and its motion is unbounded. The T-trajectory ionizes directly, i.e., it does not experience rescattering. This corresponds to a direct ionization. Indeed, in Fig.~\ref{fig:RSC_probability}, we observe that for $\xi > \xi_c$, the probability of Rydberg state creation and Coulomb-driven recollisions decreases significantly for increasing ellipticity. 
\par
Hence, for $\xi > \xi_c$, the GC motion is unbounded and the electron is driven to the detector. The initial condition of the GC of the T-trajectory is determined by combining Eqs.~\eqref{eq:initial_conditions_guiding_center_coordinates} and~\eqref{eq:most_probable_initial_momentum}, and reads
\begin{eqnarray*}
\mathbf{R}_{g,0} &=& \hat{\mathbf{x}} \dfrac{E_0}{\omega^2 \sqrt{\xi^2+1}} \cosh \tau , \\
\mathbf{P}_{g,0} &=& \hat{\mathbf{y}} \dfrac{\xi E_0}{\omega \sqrt{\xi^2+1}} \dfrac{\sinh \tau}{\tau} . 
\end{eqnarray*} 
Since Hamiltonian~\eqref{eq:H_n} is time-independent and rotationally invariant, the GC energy $\mathcal{E}_T$ and angular momentum $\ell_T = \hat{\mathbf{z}} \cdot \mathbf{R}_{g,0} \times \mathbf{P}_{g,0}$ of the T-trajectory are conserved and given by
\begin{eqnarray*}
\mathcal{E}_T &=& \dfrac{\xi^2 E_0^2}{2 \omega^2 (\xi^2+1)} \dfrac{\sinh^2 \tau}{\tau^2} - \dfrac{\omega^2 \sqrt{\xi^2+1}}{E_0 \cosh \tau} , \\
\ell_T &=& \dfrac{\xi E_0^2}{\omega^3 (\xi^2+1)} \dfrac{\sinh 2 \tau}{2 \tau} . 
\end{eqnarray*}
When the electric field is turned off, we assume that the final momentum of the T-trajectory and the final momentum of its GC are equal, with $P_x = \sqrt{2\mathcal{E}_T} \cos \Theta$ and $P_y = \sqrt{2\mathcal{E}_T} \sin \Theta$, where its scattering angle is given by $\Theta = \pi/2 + \sin^{-1} (2 \mathcal{E}_T \ell_T^2 + 1)^{-1/2}$. As a consequence,
\begin{subequations} 
\label{eq:Final_momenta_guiding_center}
\begin{eqnarray}
P_x &=& - \sqrt{2 \mathcal{E}_T} (2 \mathcal{E}_T \ell_T^2 + 1)^{-1/2} , \\
P_y &=& \sqrt{2 \mathcal{E}_T} \left[ 1 - (2 \mathcal{E}_T \ell_T^2 + 1)^{-1}\right]^{1/2} .
\end{eqnarray} 
\end{subequations}
Equations~\eqref{eq:Final_momenta_guiding_center} are used to compute $P_x$ and $P_y$ of the GC  throughout the article. In the PMDs, we recall that the bifurcation in $P_x$ signals the appearance of Coulomb asymmetry as a function of the ellipticity, while the bifurcation in $P_y$ shows the breakdown of Coulomb focusing as a function of the ellipticity. We observe that Coulomb asymmetry appears at the same time as Coulomb focusing begins to recede. Close to the bifurcation, for $\xi \approx \xi_c$ and using $\tau \approx \sinh^{-1} \gamma$, one has
\begin{subequations}
\label{eq:exponents_bifurcation_guiding_center}
\begin{eqnarray}
\mathcal{E}_T &\approx& (\xi - \xi_c) 4 \mathrm{U}_p \xi_c \gamma^2 / (\sinh^{-1} \gamma)^2 , \\
P_x &\approx& - (\xi - \xi_c)^{1/2}  \sqrt{2 \xi_c} (E_0/\omega) (\gamma/\sinh^{-1} \gamma ) , \\
P_y &\approx& (\xi-\xi_c)  2\sqrt{2} (E_0/\omega) (\gamma/\sinh^{-1} \gamma ) ,
\end{eqnarray}
\end{subequations}
where $\mathrm{U}_p = E_0^2/4\omega^2$ is the ponderomotive energy (see Ref.~\citep{Dubois2018} for more details). As a consequence, the critical exponents of the bifurcation predicted by the GC model for $P_x$ and $P_y$ are $0.5$ and $1$, respectively, i.e., $P_x \sim (\xi-\xi_c)^{1/2}$ and $P_y \sim (\xi - \xi_c)$. We observe that close to the bifurcation and for increasing ellipticity, the Coulomb asymmetry measured by the bifurcation in $P_x$ increases faster than the breakdown of Coulomb focusing measured by the bifurcation in $P_y$.

\subsubsection{Comparison with experiments \label{sec:bifurcation_experiments}}

\begin{figure}
\centering
\includegraphics[width=0.5\textwidth]{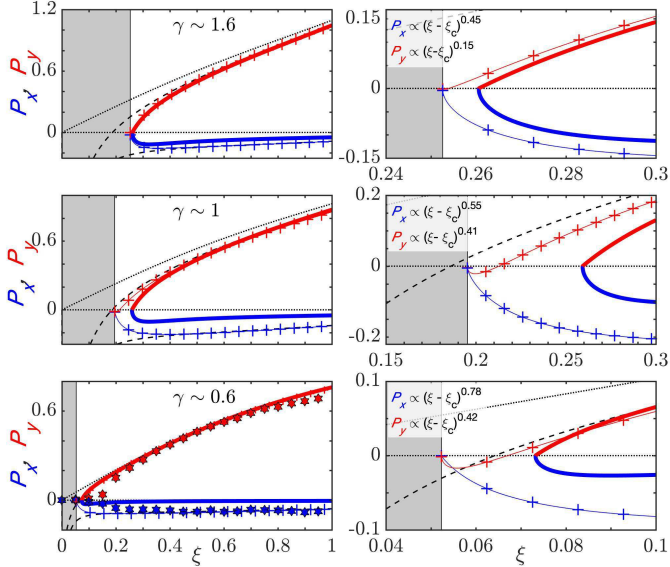}
\caption{Final momentum of the T-trajectory $\mathbf{P} = P_x \hat{\mathbf{x}} + P_y \hat{\mathbf{y}}$ as a function of the ellipticity $\xi$. Top panels: $I = 8 \times 10^{13} \; \mathrm{W}\cdot \mathrm{cm}^{-2}$, $\mathrm{He}$ ($I_p = 0.9\; {\rm a.u.}$) and $\gamma \sim 1.6$. Middle panels: $I = 1.2 \times 10^{14} \; \mathrm{W}\cdot \mathrm{cm}^{-2}$, $\mathrm{Ar}$ ($I_p = 0.58\; {\rm a.u.}$) and $\gamma \sim 1$. Bottom panels: $I = 8 \times 10^{14} \; \mathrm{W}\cdot \mathrm{cm}^{-2}$, $\mathrm{He}$ ($I_p = 0.9\; {\rm a.u.}$) and $\gamma \sim 0.6$. The hexagrams are the experimental data reproduced from Ref.~\citep{Landsman2013}. In all panels: the dotted and dashed black lines are the T-trajectory of the SFA and the CCSFA, respectively. The thin (with crosses) and solid curves are the T-trajectory of the reference Hamiltonian~\eqref{eq:Hamiltonian_electron} and the GC model~\eqref{eq:H_n}, respectively. The components of the final momentum of the T-trajectory $P_x$ and $P_y$ are depicted in blue and red, respectively. The critical ellipticity $\xi_c$ is at the intersection between the grey and white regions and corresponds to the largest ellipticity for which the T-trajectory of the reference Hamiltonian~\eqref{eq:Hamiltonian_electron} is negative. The right panels are zooms of the left panels in the neighborhood of the critical ellipticity. We indicate the scaling of $\mathbf{P}$ of the reference model~\eqref{eq:Hamiltonian_electron} in the neighborhood of the bifurcation. Momenta are scaled by $E_0/\omega$.}
\label{fig:Bifurcations_2}
\end{figure}

In Fig.~\ref{fig:Bifurcations_2}, we show the final momentum of the T-trajectory $\mathbf{P}$ as a function of the ellipticity $\xi$ computed using the SFA (dotted lines), the CCSFA from Eq.~\eqref{eq:CCSFA_estimations} (dashed lines), the reference Hamiltonian~\eqref{eq:Hamiltonian_electron} (crosses) and the GC from Eqs.~\eqref{eq:Final_momenta_guiding_center} (solid lines). The T-trajectory final momentum of the reference Hamiltonian~\eqref{eq:Hamiltonian_electron} is not depicted if it is trapped in a Rydberg state or undergoes rescattering. In the lower-left panel, the hexagrams are the experimental data of $\mathbf{P}$ reproduced from Ref.~\citep{Landsman2013}.
\par
For $I = 8\times 10^{13} \; \mathrm{W}\cdot \mathrm{cm}^{-2}$, $\mathrm{He}$ ($I_p = 0.9\; {\rm a.u.}$) and $\gamma \sim 0.6$ (top panels of Fig.~\ref{fig:Bifurcations_2}), the T-trajectory of the reference Hamiltonian~\eqref{eq:Hamiltonian_electron} corresponds to a direct ionization at the critical ellipticity $\xi_c \approx 0.25$, and reaches the detector without undergoing rescattering for $\xi > \xi_c$. The critical ellipticity is in agreement with the prediction $\xi_c \approx 0.26$ of Eq.~\eqref{eq:critical_ellipticity}. On the left panel, we observe a good agreement between the T-trajectory final momentum $\mathbf{P}$ of the reference Hamiltonian~\eqref{eq:Hamiltonian_electron} (thin curves with crosses) and that of the GC model (thick solid curves) for the entire range of ellipticities $\xi > \xi_c$.
\par
For $I = 1.2\times 10^{14} \; \mathrm{W}\cdot \mathrm{cm}^{-2}$, $\mathrm{Ar}$ ($I_p = 0.58$) and $\gamma \sim 1$ (middle panels of Fig.~\ref{fig:Bifurcations_2}), the T-trajectory of the reference Hamiltonian~\eqref{eq:Hamiltonian_electron} becomes a direct ionization at $\xi_c \approx 0.19$ while the GC prediction [see Eq.~\eqref{eq:critical_ellipticity}] is $\xi_c \approx 0.24$. There is a small disagreement between the critical ellipticity of the reference model and the prediction of Eq.~\eqref{eq:critical_ellipticity}. However, there is a good agreement of the GC critical ellipticity with the experimental measurements of Ref.~\citep{Li2017} of $\xi_c \approx 0.24$ as observed in Fig.~\ref{fig:Bifurcations_1}. Furthermore, there is a good agreement between the T-trajectory final momentum $\mathbf{P}$ of the reference Hamiltonian~\eqref{eq:Hamiltonian_electron} (thin curves with crosses) and that of the GC model (thick solid curves) for $\xi \gtrsim 0.3$. However, we observe a small disagreement between $P_x$ of the reference Hamiltonian~\eqref{eq:Hamiltonian_electron} (thin curves with crosses) and the GC prediction for all ellipticities. This discrepancy is related to the observations made in Fig.~\ref{fig:Px_Py_intensity} and whose origin is discussed below.
\par
For $I = 8\times 10^{14} \; \mathrm{W}\cdot \mathrm{cm}^{-2}$, $\mathrm{He}$ ($I_p = 0.9$) and $\gamma \sim 0.6$ (lower panels of Fig.~\ref{fig:Bifurcations_2}), the T-trajectory of the reference Hamiltonian~\eqref{eq:Hamiltonian_electron} becomes a direct electron at $\xi_c \approx 0.05$. The critical ellipticity is in agreement with the prediction $\xi_c \approx 0.07$ of Eq.~\eqref{eq:critical_ellipticity}. In addition, these values agree well with the critical ellipticity $\xi_c \approx 0.08$ of the experiments~\citep{Landsman2013} (hexagrams). There is again a good agreement between the T-trajectory final momentum $\mathbf{P}$ of the reference Hamiltonian~\eqref{eq:Hamiltonian_electron} (thin curves with crosses) and that of the GC model (thick solid curves) for $\xi \gtrsim 0.1$. However, we observe a disagreement between $P_x$ of the reference Hamiltonian~\eqref{eq:Hamiltonian_electron} (thin curves with crosses) and the GC prediction in the entire ellipticity range. We notice that for decreasing Keldysh parameters, the disagreement between $P_x$ of the reference Hamiltonian~\eqref{eq:Hamiltonian_electron} and the GC model increases, as observed in the lower panel of Fig.~\ref{fig:Px_Py_intensity}. 
\par
On the right panels of Fig.~\ref{fig:Bifurcations_2}, we observe a good agreement between the exponents of $P_x$ of the reference Hamiltonian~\eqref{eq:Hamiltonian_electron} at the bifurcation and the prediction $0.5$ of Eq.~\eqref{eq:exponents_bifurcation_guiding_center}. However, the exponent of $P_y$ at the bifurcation is much smaller than the exponent $1$ predicted by Eq.~\eqref{eq:exponents_bifurcation_guiding_center}. 
\par
In the left panels of Fig.~\ref{fig:Bifurcations_2}, we observe excellent agreement between the T-trajectory final momentum of the reference Hamiltonian~\eqref{eq:Hamiltonian_electron} (thin curves with crosses) and that of the CCSFA (dashed curves) after the bifurcation when the electron final energy is large.

\subsubsection{T-trajectory analysis \label{sec:T_trajectory_analyses}}

\begin{figure}
\centering
\includegraphics[width=0.5\textwidth]{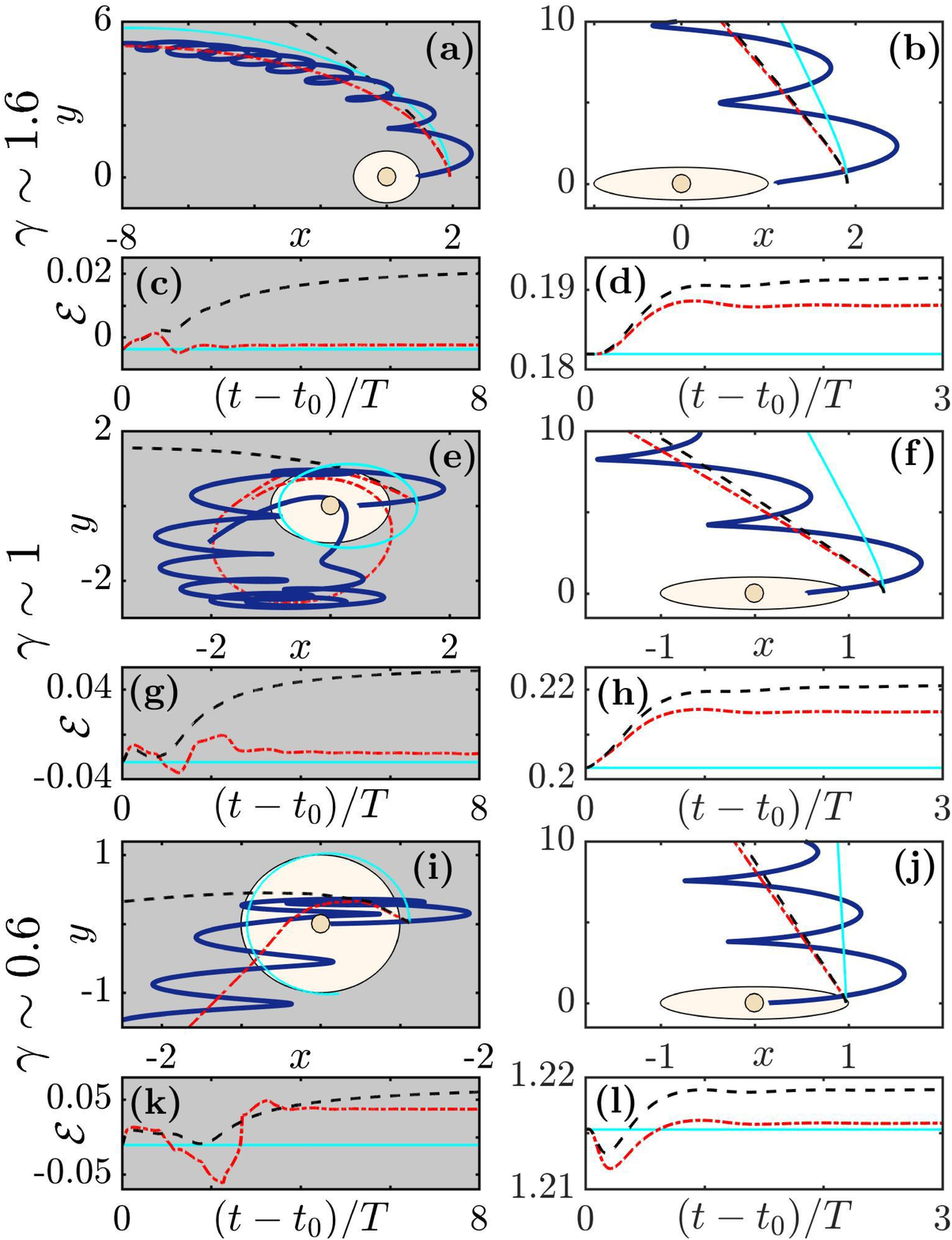}
\caption{(a,b,e,f,i,j) T-trajectory in the polarization plane $(x,y)$. The thick dark blue curves are the T-trajectory of the reference Hamiltonian. The red dash-dotted, cyan solid and dashed black curves are the T-trajectory of Hamiltonian~\eqref{eq:guiding_center_model}, the GC model~\eqref{eq:H_n} and the CCSFA~\eqref{eq:CCSFA_estimations}, respectively. (c,d,g,h,k,l) Energy~\eqref{eq:H_n} as a function of $(t-t_0)/T$, with $t_0 = T/2$, associated with each model. Right panels: $\xi = 0.7$. (a,c), (e,g) and (i,k) (the grey background panels are those for $\xi < \xi_c$) $\xi = 0.25$, $\xi = 0.15$ and $\xi = 0.05$, respectively. (a--d) are for $I = 8\times 10^{13} \; \mathrm{W}\cdot \mathrm{cm}^{-2}$, $\mathrm{He}$ ($I_p = 0.9\; {\rm a.u.}$) and $\gamma \sim 1.6$ (same parameters as the top panels of Fig.~\ref{fig:Bifurcations_2}). (e--h) are for $I = 1.2\times 10^{14} \; \mathrm{W}\cdot \mathrm{cm}^{-2}$, $\mathrm{Ar}$ ($I_p = 0.58\; {\rm a.u.}$) and $\gamma \sim 1$ (same parameters as the middle panels of Fig.~\ref{fig:Bifurcations_2}). (i--l) are for $I = 8\times 10^{14} \; \mathrm{W}\cdot \mathrm{cm}^{-2}$, $\mathrm{He}$ ($I_p = 0.9\; {\rm a.u.}$) and $\gamma \sim 0.6$ (same parameters as the lower panels of Fig.~\ref{fig:Bifurcations_2}). The dots indicate the origin, and the circles $|\mathbf{r}| = E_0/\omega^2$. The distances are scaled by $E_0/\omega^2$.}
\label{fig:T_trajectory_analyses}
\end{figure}

Here, we show that the origin of the disagreements between the T-trajectory of the reference Hamiltonian~\eqref{eq:Hamiltonian_electron} and the GC T-trajectory --the disagreement of $P_x$ for small Keldysh parameters, or the disagreement with the critical exponents of $P_y$ in the neighborhood of the bifurcation-- are related to an underestimate of the Coulomb interaction by the GC model for a short time after ionization. In contrast, we show that the CCSFA agrees well with the solution of the reference Hamiltonian~\eqref{eq:Hamiltonian_electron} for $\xi \gg \xi_c$ while it cannot capture correctly the phenomena related to the bifurcation.
\par
In Fig.~\ref{fig:T_trajectory_analyses}, the red dash-dotted, cyan solid and black dashed curves are the T-trajectory of Hamiltonian~\eqref{eq:guiding_center_model}, the GC model~\eqref{eq:H_n}, and the CCSFA given by Eqs.~\eqref{eq:CCSFA_estimations}, respectively. The thick dark blue curves are the T-trajectory of the reference Hamiltonian~\eqref{eq:Hamiltonian_electron}. Associated with each trajectory, we also show the GC energy, for each model, as a function of time per laser cycle $t/T$. The GC energy for each model consists substituting the solution $(\mathbf{r}_{g} (t) , \mathbf{p}_g (t))$ for each model in the GC Hamiltonian~\eqref{eq:H_n}, i.e., $\bar{H} (\mathbf{r}_g (t) , \mathbf{p}_g (t) )$. Where the GC energy of the reference model is conserved, the GC model (whose GC energy is conserved) is valid~\citep{Dubois2018_PRE}.
\par
For $\gamma \sim 1.6$ (see Fig.~\ref{fig:T_trajectory_analyses}a--d), the electron ionizes far from the ionic core ($|\mathbf{r}_0| \sim E_0/\omega^2$). 
For $\xi = 0.25$ and $\xi = 0.7$, respectively, we see the variations of the GC energy of the T-trajectory of Hamiltonian~\eqref{eq:guiding_center_model} (dash-dotted curve) are small, a signature of the validity of the GC model and an absence of rescattering. When the GC energy of Hamiltonian~\eqref{eq:guiding_center_model} becomes constant, it is only about $0.02 \; \mathrm{a.u.}$ above the GC model prediction. As a consequence, we observe a good agreement between the trajectories of Hamiltonian~\eqref{eq:guiding_center_model} and the GC model trajectories in Fig.~\ref{fig:T_trajectory_analyses}a and Fig.~\ref{fig:T_trajectory_analyses}b. 
In particular, at $\xi = 0.25$, we observe the T-trajectory of Hamiltonian~\eqref{eq:guiding_center_model} is trapped in a Rydberg state, a feature which is reproduced by the GC model (cyan solid curve), but not well reproduced by the CCSFA (dashed black curve). Indeed, the Coulomb interaction remains significant for a long time after ionization during Rydberg state creation, and the conditions for the validity of the CCSFA are not met.
\par
For $\gamma \sim 1$ (see Fig.~\ref{fig:T_trajectory_analyses}e--h), the electron ionizes closer to the ionic core ($|\mathbf{r}_0| \sim 0.4 E_0/\omega^2$). 
For $\xi = 0.15$, the electron T-trajectory of Hamiltonian~\eqref{eq:guiding_center_model} (dash-dotted curve) and the GC model are trapped in Rydberg states. However, there is a large discrepancy between the trajectories. Indeed, in Fig.~\ref{fig:T_trajectory_analyses}g, we observe that the dash-dotted red curve varies after ionization, indicating that the electron rescatters for a short time after ionization~\citep{Dubois2018_PRE}. 
For $\xi = 0.7$, the same happens in terms of energy (see Fig.~\ref{fig:T_trajectory_analyses}h), and we see that the GC trajectory does not agree well with the T-trajectory of Hamiltonian~\eqref{eq:guiding_center_model} (see Fig.~\ref{fig:T_trajectory_analyses}f). When the energy of Hamiltonian~\eqref{eq:guiding_center_model} becomes constant at $t \approx T/2$, it is larger than the GC energy prediction of $0.15 \; \mathrm{a.u.}$ In Fig.~\ref{fig:T_trajectory_analyses}e or Fig.~\ref{fig:T_trajectory_analyses}f, we observe that after ionization, the initial electron distance from the core is $|\mathbf{r}_0| \sim 0.4 E_0/\omega^2$, while the GC initially at a distance $|\mathbf{r}_{g,0} | \sim 1.4 E_0/\omega^2$ from core. Since in the GC model, the Coulomb interaction is evaluated at the GC position only, when the electron is closer to the core than predicted by the GC model, as is the case after ionization for $\gamma \lesssim 1.6$, the Coulomb interaction is underestimated in the GC model: the closer the electron to the ionic core, the more underestimated the Coulomb interaction. 
\par
For $\gamma \sim 0.6$ (see Figs.~\ref{fig:T_trajectory_analyses}i--l), the electron ionizes even closer to the ionic core ($|\mathbf{r}_0| \sim 0.15 E_0/\omega^2$). 
For $\xi = 0.05$ and $\xi = 0.07$, there are also discrepancies between the cyan and red curves. We observe that the energy of the T-trajectory of Hamiltonian~\eqref{eq:guiding_center_model} (red dash-dotted curve in Figs.~\ref{fig:T_trajectory_analyses}k and~\ref{fig:T_trajectory_analyses}l) varies a lot for a short time after ionization (about $0.2T$). Here again, the electron rescatters after ionization. 
In Fig.~\ref{fig:T_trajectory_analyses}l, when the red dash-dotted curve becomes constant, the energy is above the GC prediction only by $0.02 \; \mathrm{a.u.}$ However, this agreement is only coincidental since the T-trajectories of Hamiltonian~\eqref{eq:guiding_center_model} and of the GC disagree significantly due to the increase in energy of the rescattering. We observe that this increase in energy after ionization is well captured by the CCSFA. 
\par
In each panel, we observe an excellent agreement between the CCSFA and the T-trajectory of Hamiltonian~\eqref{eq:guiding_center_model} for a short time after ionization, i.e., $0 < t - t_0 \lesssim T$, when the hypotheses of the CCSFA are met. This method is effective for short-time dynamics or phenomena~\citep{Kamor2014,Kastner2012,Danek2018}. This agreement persists for longer times if the electron leaves quickly the ionic core region like in Ref.~\citep{Goreslavski2004} or for large ellipticity (see Sec.~\ref{sec:comparison_models_intensity}), i.e., if its drift momentum is initially large. 

\section*{Conclusions}
In this article, we have investigated the role of the Coulomb potential in atoms subjected to strong laser fields. We have considered three reduced models of the reference Hamiltonian~\eqref{eq:Hamiltonian_electron} to do so, namely the SFA [Eqs.~\eqref{eq:SFA_solutions}], the CCSFA [Eqs.~\eqref{eq:CCSFA_estimations}], and the GC model [Eqs.~\eqref{eq:H_n}]. The analysis of these three reduced models allowed us to shed light on the manifestation of the Coulomb potential in various ionization processes. In the SFA, there are two types of trajectories: subcycle recollisions and direct ionizations. However, even when the intensity is very large, i.e., when the conditions of the SFA are met, the Coulomb interaction still makes its presence known for long time scale phenomena. In particular, even at very high intensities, the Coulomb asymmetry persists as seen in Fig.~\ref{fig:Px_Py_intensity} and discussed in Sec.~\ref{sec:comparison_models_intensity}. The Coulomb interaction brings with it a variety of additional types of trajectories, such as Coulomb-driven recollisions and Rydberg states. We have shown in Sec.~\ref{sec:Coulomb_driven_Rydberg_states} that these two processes are intimately related, and can be interpreted and predicted by the GC model.
\par
During step (ii) of the recollision scenario, we have shown that the electron oscillates around the GC trajectory. In phase space, the GC trajectory lies on a curve of constant energy $\mathcal{E} = \bar{H} (\bar{\mathbf{r}}_g , \bar{\mathbf{p}}_{g})$. If $\mathcal{E}>0$, the GC motion is unbounded. In this case, it is likely the electron recollides if its GC angular momentum is near zero and its initial radial momentum is negative (like in Fig.~\ref{fig:tunneling_exit}b). Otherwise, the electron ionizes directly without recollision (like in Fig.~\ref{fig:tunneling_exit}c). If $\mathcal{E}<0$, the GC motion is bounded. In this case, there exists at least one time at which the electron turns back towards the ionic core far from the origin. Then, the electron returns to the ionic core before the laser field is turned off and recollides, or not. If the electron does recollide (like in Figs.~\ref{fig:tunneling_exit}d and~\ref{fig:laser_pulse_duration}d), the GC energy jumps to a new energy level, as described in Ref.~\citep{Dubois2018_PRE}. If the laser field is turned off before the electron recollides (as in Figs.~\ref{fig:tunneling_exit}e and~\ref{fig:laser_pulse_duration}e), the electron ends up on a Rydberg state. 
\par
During step (iii), the electron is rescattered by the ionic core. The GC model does not capture the rescattering effects close to the ionic core but the CCSFA can since it is a rather short time scale phenomenon~\citep{Kastner2012,Danek2018}. After a close encounter with the ionic core, the GC energy jumps to another energy level. As observed in Fig.~\ref{fig:T_trajectory_analyses}, the variations of energy of the reference model can be well described by the CCSFA for short time scales. After rescattering, the electron potentially ionizes if its GC energy becomes positive (such as in Fig.~\ref{fig:tunneling_exit}d). Therefore, the CCSFA and the GC models are clearly complementary. The CCSFA is adapted for describing short time scale processes such as rescattering while the GC model is more suited for describing long time scale processes such as Coulomb-driven recollisions and the creation of Rydberg states.

\section*{Acknowledgments}
We thank Jan-Michael Rost, Ulf Saalmann, and Cornelia Hoffman for helpful discussions, and Ursula Keller, Yunquan Liu and their groups for sharing their experimental results with us. The project leading to this research has received funding from the European Union's Horizon 2020 research and innovation program under the Marie
Sk\l{}odowska-Curie grant agreement No. 734557. T.U. and S.A.B. acknowledges funding from the NSF (Grant No.
PHY1602823).

\appendix
\section{Ionization rate \label{app_sec:ionization_rate}}
Throughout this article, we use the nonadiabatic ionization rate given by the Perelomov-Popov-Terent'ev~\citep{PerelomovII1967} formulas, rewritten in a different form in Ref.~\citep{Mur2001}. We denote $\gamma_0 (t_0) = \omega \sqrt{2 I_p} / |\mathbf{E}(t_0)|$. The initial position of the electron is parametrized by the ionization time $t_0$ and its initial momentum is written as $\mathbf{p}_0 = p_{\parallel} \hat{\mathbf{n}}(t_0) + p_{\perp} \hat{\mathbf{n}} (t_0) + p_{z,0} \hat{\mathbf{z}}$ for a polarization plane $(\hat{\mathbf{x}}, \hat{\mathbf{y}})$. The PPT ionization rate~\citep{PerelomovII1967} reads
\begin{eqnarray}
\label{app_eq:PPT_ionization_rate}
&& W (t_0, \mathbf{p}_0 ) \propto \dfrac{h (\gamma_0 (t_0), \xi)}{|\mathbf{E}(t_0)|} \exp \left[ - \dfrac{2 I_p}{\omega} g (\gamma_0 (t_0), \xi)\right]  \nonumber\\
\times&& \exp \left\lbrace - \dfrac{1}{\omega} \left[ c_{\parallel} p_{\parallel}^2 + c_{\perp} \left( p_{\perp} - P_{\perp} \right)^2 + c_{z} p_{z,0}^2 \right] \right\rbrace , 
\end{eqnarray}
where the functions $g$ and $h$ are
\begin{eqnarray*}
g (\gamma_0, \xi) &=& \left( 1 + \dfrac{1+\xi^2}{2\gamma_0^2} \right) \tau_0 \\
					&&- (1 - \xi^2) \dfrac{\sinh 2 \tau_0}{4\gamma_0^2} - \xi^2 \dfrac{\sinh^2 \tau_0}{\gamma_0^2 \tau_0} , \\
h(\gamma_0,\xi) &=& \dfrac{2 \sigma \gamma_0}{\sinh 2 \tau_0} ,
\end{eqnarray*}
with the notation
\begin{equation*}
\sigma = \left( 1 - \xi^2 + \xi^2 \dfrac{\tanh \tau_0}{\tau_0} \right)^{-1} .
\end{equation*} 
The coefficients $c_{\parallel}$ $c_{\perp}$, and $c_z$, which are inversely proportional to the square of the standard deviation of the distribution along the longitudinal and transverse momentum, are given by
\begin{eqnarray*}
c_{\parallel} &=& \tau_0 - \sigma \tanh \tau_0 , \\
c_{\perp} &=& \tau_0 + \sigma \xi^2 \dfrac{\left( \tau_0 - \tanh \tau_0 \right)^2}{\tau_0^2 \tanh\tau_0} , \\
c_{z} &=& \tau_0 .
\end{eqnarray*}
The coefficients satisfy $c_{\parallel} > c_{\perp}$, implying that the distributions are more spread out along the transverse direction than along the longitudinal direction. The most probable initial transverse momentum $p_{\perp,0}^{\max}$ is 
\begin{equation*}
p_{\perp,0}^{\max} = \dfrac{\xi E_0}{\omega \sqrt{\xi^2+1}} \left( 1 - \dfrac{\sinh \tau_0}{\tau_0} \right) ,   
\end{equation*}
for a transverse unitary vector defined as $\hat{\mathbf{n}}_{\perp} (t_0) = - [ \hat{\mathbf{n}}_{\parallel} (t_0) \cdot \hat{\mathbf{y}} ] \hat{\mathbf{x}} + [ \hat{\mathbf{n}}_{\parallel} (t_0) \cdot \hat{\mathbf{x}} ] \hat{\mathbf{y}}$. Throughout the article, we take $p_{z,0} = 0$.

\section{Final momentum of the electron in the GC model \label{app_sec:final_momentum_guiding_center}}

In the GC model given in Eq.~\eqref{eq:H_n}, the energy $\mathcal{E}$ and the angular momentum $\ell = \mathbf{r}_g \times \mathbf{p}_g \cdot \hat{\mathbf{z}}$ are conserved. The model is accurate far from the core, and as a consequence we assume that $V(\mathbf{r}_g) = -1/|\mathbf{r}_g|$. The Hamiltonian reads
\begin{equation*}
\bar{H}_g (\mathbf{r}_g , \mathbf{p}_g) = \dfrac{|\mathbf{p}_g|^2}{2} - \dfrac{1}{|\mathbf{r}_g|} .
\end{equation*}
Using polar coordinates, the Hamiltonian becomes
\begin{equation*}
\bar{H}_g (r,\theta,p_{r} ) = \dfrac{p_r^2}{2} + \dfrac{\ell^2}{2 r^2} - \dfrac{1}{r} ,
\end{equation*}
where the energy $\mathcal{E} = \bar{H}_g (r,\theta,p_{r})$. We denote $(r_0, \theta_0, p_{r,0} )$ the initial conditions in the polar coordinates. They are determined from the inverse transformation $p_r = \mathbf{p}_g \cdot \mathbf{r}_g/|\mathbf{r}_g|$, $\ell = \mathbf{r}_g \times \mathbf{p}_g \cdot \hat{\mathbf{z}}$, $r= |\mathbf{r}_g|$ and $\cos \theta = \hat{\mathbf{x}} \cdot \mathbf{r}/|\mathbf{r}|$, $\sin \theta = \hat{\mathbf{y}} \cdot \mathbf{r}/|\mathbf{r}|$.
\par
If $\mathcal{E}<0$, the electron motion is bounded. The two turning points at which the electron radial momentum changes sign are the perihelion $r_-$ (closest distance of the orbit from the core) and the aphelion $r_+$ (largest distance of the orbit from the core) such that
\begin{equation}
\label{eq_app:perihelion_aphelion_negative_E}
r_{\pm} = \dfrac{1}{2 |\mathcal{E}|} \left( 1 \pm \sqrt{1 - 2 \ell^2 |\mathcal{E}|} \right) .
\end{equation}
\par
If $\mathcal{E}>0$, the GC trajectory is unbounded and the electron reaches the detector. The asymptotic configuration (when $r$ goes to infinity) is given by $p_{r} = \sqrt{2\mathcal{E}}$. Concerning the final scattering angle $\theta$, if $\ell = 0$, $\theta = \theta_0$ (if $p_{r,0}>0$) and $\theta_0 + \pi$ (if $p_{r,0}<0$). If $\ell \neq 0$, the final scattering angle is given by
\begin{equation}
\label{app_eq:final_scattering_angle}
\theta = \left\lbrace
\begin{array}{l @{\; \text{if} \; } l}
\theta_0 + \sin^{-1} u_0 - \sin^{-1} u_{\infty} & p_{r,0} > 0 , \\
\theta_0 + \sin^{-1} u_m - \sin^{-1} u_{\infty} & p_{r,0} = 0 , \\
\theta_0 + 2 \sin^{-1} u_m - \sin^{-1} u_0 - \sin^{-1} u_{\infty} & p_{r,0} < 0 ,
\end{array} \right. 
\end{equation}
with $u_{0} = \beta ( \ell / r_0  - 1/\ell )$, $u_{\infty} = - \beta/\ell$ and $u_m = + 1$ (resp. $-1$) if $\ell > 0$ (resp. $< 0$) with $\beta = (2 \mathcal{E} + 1/\ell^2)^{-1/2}$. Finally, the final momentum of the GC in the Cartesian coordinates is given by
\begin{equation*}
\mathbf{p} = p_r \left( \hat{\mathbf{x}} \cos \theta + \hat{\mathbf{y}} \sin\theta \right) .
\end{equation*}

\end{document}